\documentclass[11pt]{article}

\usepackage{fullpage}
\usepackage[dvips]{color}
\usepackage{xcolor}
\usepackage{latexsym,dsfont}
\usepackage[activeacute,english]{babel}
\usepackage{graphicx}
\usepackage{amsmath}
\usepackage{sidecap}
\usepackage{algorithmic}
\usepackage{algorithm}
\usepackage[round]{natbib}
\usepackage{amsthm,appendix,url,inconsolata,wrapfig,framed,ulem}

\usepackage{multirow}

\usepackage{amscd}
\usepackage{amsfonts}
\usepackage[tableposition=top]{caption}
\usepackage{wrapfig}
\usepackage{ifthen}
\usepackage[utf8]{inputenc}
\usepackage{mathtools}
\newtheorem{theorem}{Theorem}
\newtheorem{prop}[theorem]{Proposition}

\theoremstyle{definition}
\newtheorem{defn}[theorem]{Definition}
\newtheorem{examp}[theorem]{Example}

\begin{document}
\title{Enumeration of binary trees  \\ compatible with a perfect phylogeny}
%Alternative: Enumeration of tree shapes compatible with a perfect phylogeny
%Alternative: On the exact number of trees compatible with molecular sequence data
%Alternative 1: Enumeration of trees conditioned on the data
%Alternative 2: Enumeration of trees compatible with a perfect phylogeny
%Alternative 3: On the number of ranked tree shapes compatible with molecular sequence data
%Alternative 4: Trees compatible with a perfect phylogeny
\author{Julia A. Palacios\thanks{Department of Statistics and Department of Biomedical Data Science, Stanford University, Stanford, CA, USA. Corresponding author: juliapr@stanford.edu} \\ Anand Bhaskar\thanks{Department of Genetics, Stanford University, Stanford University, Stanford, CA, USA.} \\ Filippo Disanto\thanks{Department of Mathematics, University of Pisa, Pisa, Italy.} \\ Noah A. Rosenberg\thanks{Department of Biology, Stanford University, Stanford, CA, USA.}}
%\texttt{$^\star$Corresponding author: juliapr@stanford.edu}
\date{\today}
\maketitle

\begin{abstract}
Evolutionary models used for describing molecular sequence variation suppose that at a non-recombining genomic segment, sequences share ancestry that can be represented as a genealogy---a rooted, binary, timed tree, with tips corresponding to individual sequences. Under the infinitely-many-sites mutation model, mutations are randomly superimposed along the branches of the genealogy, so that every mutation occurs at a chromosomal site that has not previously mutated; if a mutation occurs at an interior branch, then all individuals descending from that branch carry the mutation. The implication is that observed patterns of molecular variation from this model impose combinatorial constraints on the hidden state space of genealogies. In particular, observed molecular variation can be represented in the form of a perfect phylogeny, a tree structure that fully encodes the mutational differences among sequences.
%Under this model, observed molecular variation can be represented as a perfect phylogeny (also known as gene tree) imposing constraints in the topology of underlying coalescent tree. 
For a sample of $n$ sequences, a perfect phylogeny might not possess $n$ distinct leaves, and hence might be compatible with many possible binary tree structures that could describe the evolutionary relationships among the $n$ sequences. Here, we investigate enumerative properties of the set of binary ranked and unranked tree shapes that are compatible with a perfect phylogeny, and hence, the binary ranked and unranked tree shapes conditioned on an observed pattern of mutations under the infinitely-many-sites mutation model. We provide a recursive enumeration of these shapes. We consider both perfect phylogenies that can be represented as binary and those that are multifurcating. The results have implications for computational aspects of the statistical inference of evolutionary parameters that underlie sets of molecular sequences.   

%describe the cardinality of the state space of different resolutions of the Kingman-Tajima coalescent conditioned on the observed perfect phylogeny and the infinite sites model of mutations. Coalescent-based inference of population genetic parameters from molecular sequence data rely on integration over the hidden state space of coalescent trees. Our results dictate whether inference of such parameters is computationally feasible or not for a given dataset. 

\end{abstract}

%%Places for publication: TPB, Journal of Mathematical Biology (like Amandine's), Annals of applied probability

\section{Introduction}

Coalescent and mutation models are used in population genetics  to estimate evolutionary parameters from samples of molecular sequences \citep{marjoram2006modern}. The central idea is that observed molecular variation is the result of a process of mutation along the branches of the genealogy of the sample. This genealogy is a timed tree that represents the ancestral relationships of the sample at a chromosomal segment. Consisting of a tree topology and its branch lengths, the genealogy is a nuisance parameter that is modeled as a realization of the coalescent process dictated by evolutionary parameters---which are in turn inferred by integrating over the space of genealogies. For large sample sizes, however, this integration is computationally challenging because the state space of tree topologies increases exponentially with the number of sampled sequences. 

%\textcolor{green}{as coalescent models have diversified to address computational challenges (Spence et al., 2018)},

Recently, a coarser coalescent model known as the \emph{Tajima coalescent} \citep{Tajima1983,veber}, coupled with the infinitely-many-sites mutation model \citep{Kimura69} has been introduced for population-genetic inference problems \citep{Palacios2019}. Whereas the standard coalescent model \citep{Kingman:1982uj} induces a probability measure on the space of ranked labeled tree topologies, the Tajima coalescent induces a probability measure on the space of ranked \emph{unlabeled} tree topologies. Removing the labels of the tips from the tree topology, as in the Tajima coalescent, reduces the cardinality of the space of tree topologies substantially, shrinking computation time in inference problems. 

Under infinitely-many-sites mutation, only a subset of tree topologies (labeled or unlabeled) are compatible with an observed data set, so that the computational complexity of inference varies among different data sets. Hence, \citet{Cappello2019} used importance sampling to approximate cardinalities of the spaces of labeled and unlabeled ranked tree shapes conditioned on a data set of molecular sequences, demonstrating a striking reduction of the cardinality of the space of ranked unlabeled tree shapes versus the labeled counterpart when conditioning on observed data with a sparse number of mutations. Here, we extend beyond the approximate work of  \citet{Cappello2019} and obtain exact results. We provide a recursive algorithm for exact computation of the cardinality of the spaces of labeled and unlabeled ranked tree shapes compatible with a sequence data set. We provide a number of other enumerative results relevant for inference of tree topologies in phylogenetics and population genetics. Python code for enumeration is available at \\ %\begin{verbatim}
{\tt https://colab.research.google.com/drive/1cAx2xyn7OtmG-F-9nxJ3CHRc7e7AjuCj?usp=sharing}. 
%\end{verbatim}

%In this work, our aim is to understand the complexity of the state space of Tajima's genealogies conditioned on the data and the infinite sites mutation model. 

\section{Preliminaries}

\subsection{Types of trees}

\begin{figure}
\begin{center}
\includegraphics[scale=0.46]{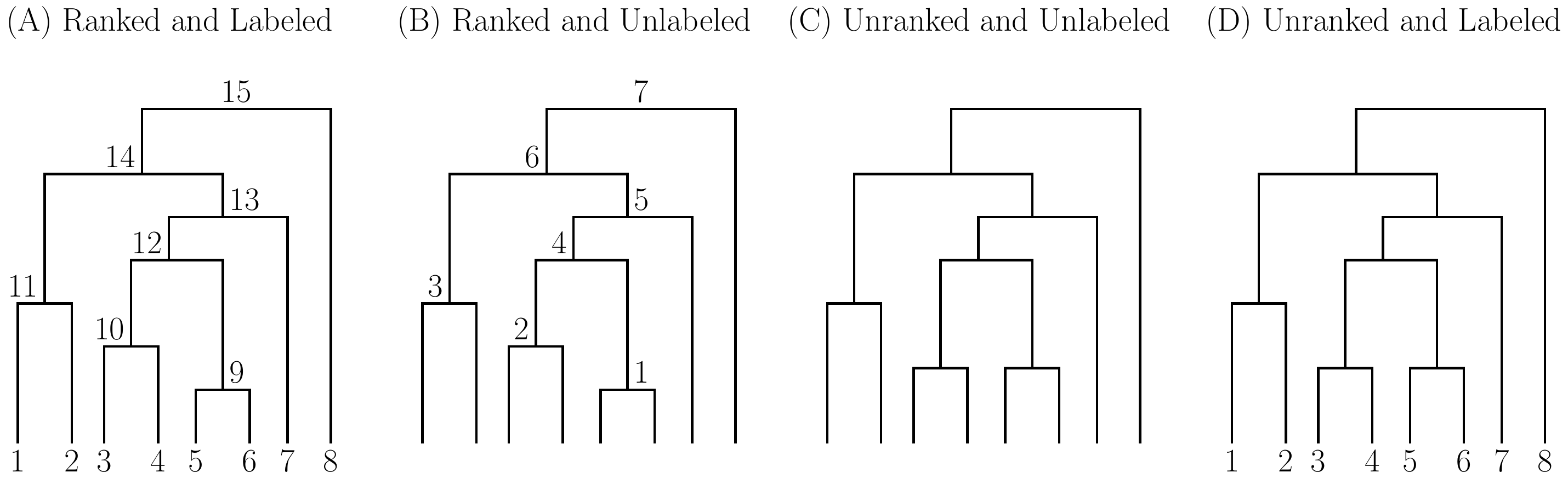}
 \end{center}
\caption{\small{\textbf{Different types of trees.}  (A) A ranked labeled tree shape. (B) A ranked unlabeled tree shape. (C) An unranked unlabeled tree shape. (D) An unranked labeled tree shape. The ranked unlabeled tree shape in (B) is obtained by discarding leaf labels from the ranked labeled tree shape in (A). The unranked labeled tree shape in (D) is obtained by discarding the sequence of internal node ranks in (A). The unranked unlabeled tree shape in (C) is obtained by discarding the sequence of internal node ranks in (B) or the leaf labels in (D).}}
\label{fig:tree_topologies}
\end{figure}

The \textbf{coalescent} is a continuous-time Markov chain with values in the space $\mathcal{P}_{n}$ of partitions of $[n]=\{1,2,\ldots,n\}$ \citep{Kingman:1982uj}. The process starts with the trivial partition of $n$ singletons, labeled $\{1\},\{2\},\ldots,\{n\}$, at time 0; at each transition, two blocks are chosen uniformly at random to merge into a single block. The process ends with a single block with label $\{1,2,\ldots,n\}$. In the standard coalescent, the holding times are exponentially distributed with rate $\binom{k}{2}$ when there are $k$ blocks. Transition probabilities for the coalescent can be factored into two independent components, a pure death process and a discrete jump chain. A full realization of the process can be represented by a timed rooted binary tree: a genealogy. The tips of the genealogy are labeled by $\{1,2,\ldots,n\}$. Figure \ref{fig:tree_topologies}A shows a realization of the jump process, a ranked labeled tree shape.

A lumping of the standard coalescent process, called the \textbf{Tajima coalescent} \citep{veber}, consists in removing the labels of the tips of the genealogy. The pure death process of the lumped process is the same as the standard coalescent. The discrete jump chain can be described as a simple urn process \citep{janson2011}. Start with an urn of $n$ balls labeled $0$; at the $i$th transition, draw two balls and return one to the urn with label $i$. The process ends when there is a single ball with label $n-1$ in the urn. A full realization of the urn process can be represented as a ranked unlabeled tree shape with internal nodes labeled by the transition index.

%%%%%%%%%%%%%%%%%%%%%%%%% Figure 1 %%%%%%%%%%%%%%%%%%%%%%%%%%%%%%%%%%
%%%%%%%%%%%%%%%%%%%%%%%%%%%%%%%%%%%%%%%%%%%%%%%%%%%%%%%%%%%%%

A \textbf{ranked labeled tree shape} of size $n$, denoted by $T^{L}_{n}$, is a rooted binary labeled tree of $n$ leaves with a total ordering for the internal nodes. Without loss of generality, we use label set $[n]$ to label the $n$ leaves. The space of ranked labeled tree shapes with $n$ leaves will be denoted by $\mathcal{T}^{L}_{n}$.  Figure \ref{fig:tree_topologies}A shows an example of a ranked labeled tree shape with $n=8$ leaves. Ranked labeled tree shapes are also known as labeled histories.

\begin{figure}
  \begin{center}
\includegraphics[scale=0.5]{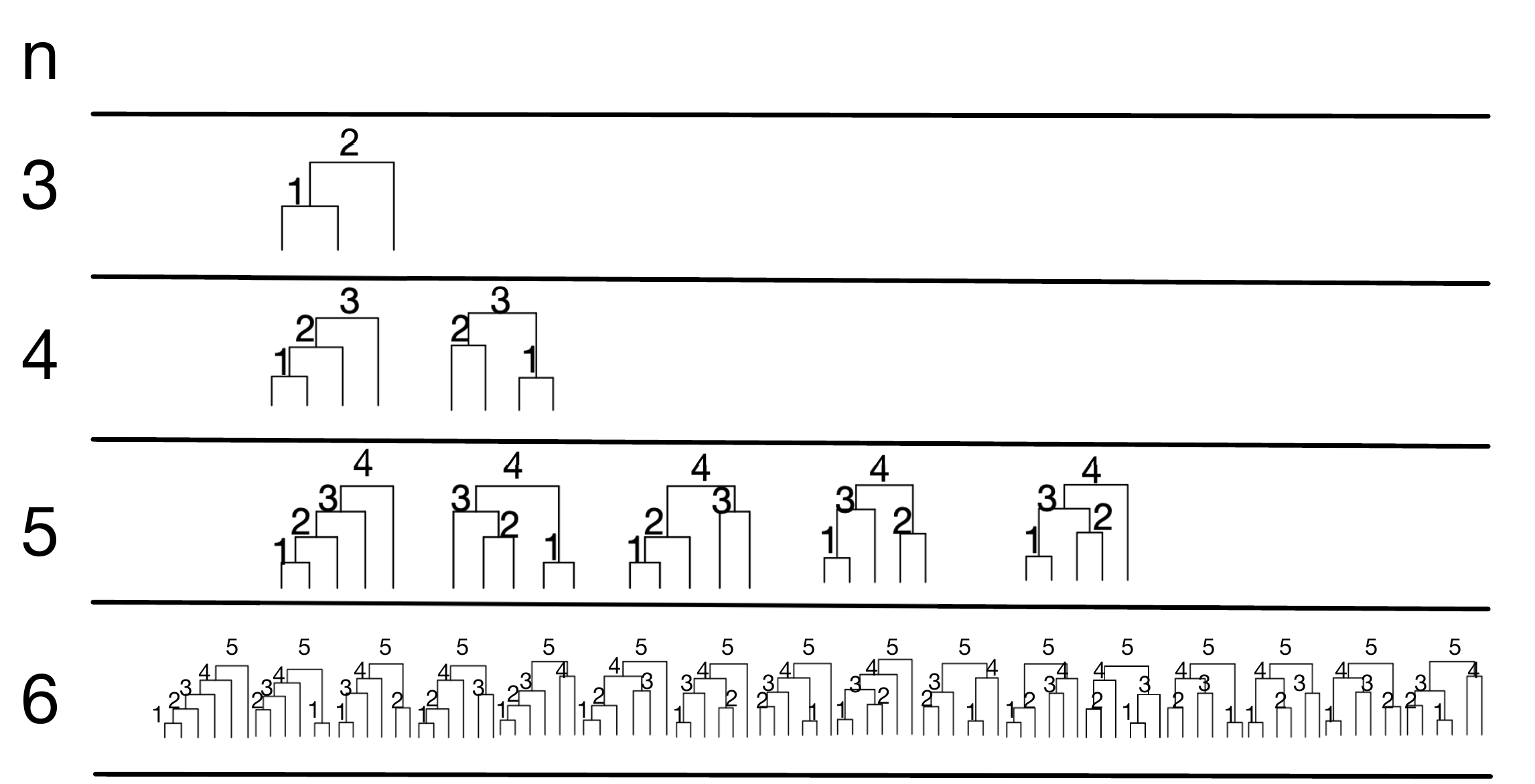}
%  % \includegraphics[width=1.0\textwidth]{Figures/tree_s.pdf}
  \end{center}
\caption{\small{\textbf{An enumeration of all possible ranked tree shapes with 3, 4, 5, and 6 leaves.} }}
\label{fig:trees1}
\end{figure}

A \textbf{ranked unlabeled tree shape} of size $n$, denoted by $T^{R}_{n}$, is a rooted binary unlabeled tree of $n$ leaves with a total ordering for the internal nodes. The space of ranked unlabeled tree shapes with $n$ leaves will be denoted by $\mathcal{T}^{R}_{n}$. Figure \ref{fig:tree_topologies}B shows an example of a ranked unlabeled tree shape with $n=8$ leaves. We will refer to a ranked unlabeled tree shape simply as a ranked tree shape; these ranked tree shapes are also known as unlabeled histories, or Tajima trees. Figure \ref{fig:trees1} shows all ranked unlabeled tree shapes with $3,4,5,$ and $6$ leaves.

%%%%%%%%%%%%%%%%%%%%%%%%%%%%%%%%%%%%%%%%%%%%%%%%%%%%%%%%%%%%

%A lumping of the Tajima coalescent, called \textbf{sized n-coalescent} \citep[unvitanged and sized n-coalescent]{veber} consists in removing the labels of the internal nodes. Again, the pure death process of the lumped process is the same as the standard coalescent and the discrete jump chain keeps track the sizes of the blocks and can be described as an urn process. Start with an urn of $n$ balls labeled 1; at the i-th transition draw two balls and return one to the urn with label the sum of the two labels of the two balls drawn. The process ends when there is a single ball with label $n$ in the urn. A full realization of the urn process can be represented as a tree shape.

An \textbf{unranked unlabeled tree shape} of size $n$, denoted by $T_{n}$, is a rooted binary unlabeled tree of $n$ leaves with unlabeled internal nodes. The space of unranked (unlabeled) tree shapes with $n$ leaves will be denoted by $\mathcal{T}_{n}$. Figure \ref{fig:tree_topologies}C shows an example of an unranked unlabeled tree shape with $n=8$ leaves. 
%We will refer to an unranked unlabeled tree shape simply as a tree shape; 
These shapes are also called unlabeled topologies or Otter trees \citep{otter1948number}.

An \textbf{unranked labeled tree shape} of size $n$, denoted by $T^X_n$, is a rooted binary labeled tree of $n$ leaves with unlabeled internal nodes. The space of unranked labeled tree shapes with $n$ leaves will be denoted by $\mathcal{T}^X_{n}$. Figure \ref{fig:tree_topologies}D shows an example of an unranked labeled tree shape with $n=8$ leaves. These tree shapes are also called labeled topologies.

%Following notation of \cite{veber}, let $\neg(T_{n})$ denote the number of events that arise from coalescing two blocks of distinct sizes, and let $f_{i,j}$ denote the number of blocks of size $j$ while there are $i$ blocks, then the probability of a ranked tree shape generated with the sized n-coalescent is $\frac{2^{\neg(T_{n})}}{(n-1)!} \prod^{n}_{i=2}f_{i-1,k_{i}}$, where $k_{i}=\min \{ k (f_{i,k}-f_{i-1,k}) \}^{n}_{k=1}$ \citep{veber}.

\subsection{Mutations on trees}

Many generative models of neutral molecular evolution assume that a process of mutations is superimposed on the genealogy as a continuous-time Markov process. In the \textbf{infinitely-many-sites mutation model}, every mutation along the branches of the tree occurs at a chromosomal site that has not previously mutated \citep{Kimura69}. Therefore, if a mutation occurs at an interior branch along the genealogy, all sequences descended from that branch carry the mutation. Because every site can mutate at most once, the sequence of mutated sites can be encoded as a binary sequence, with 0 denoting the ancestral type and 1 denoting the mutant type at any site. 

Figure \ref{fig:data0}A shows a realization of the Tajima coalescent together with a realization of mutations from the infinitely-many-sites mutation model with 5 individuals and 4 mutated sites. In what follows, we assume that we observe molecular data only as binary sequences at the tips of the tree.

\subsection{Observed binary molecular sequence data as a perfect phylogeny}\label{sec:data}

%%%%%%%%%%%%%%%%%%%%%%%% Figure 3 %%%%%%%%%%%%%%%%%%%%%%%%%%%
\begin{figure}
  \begin{center}
\includegraphics[scale=0.5]{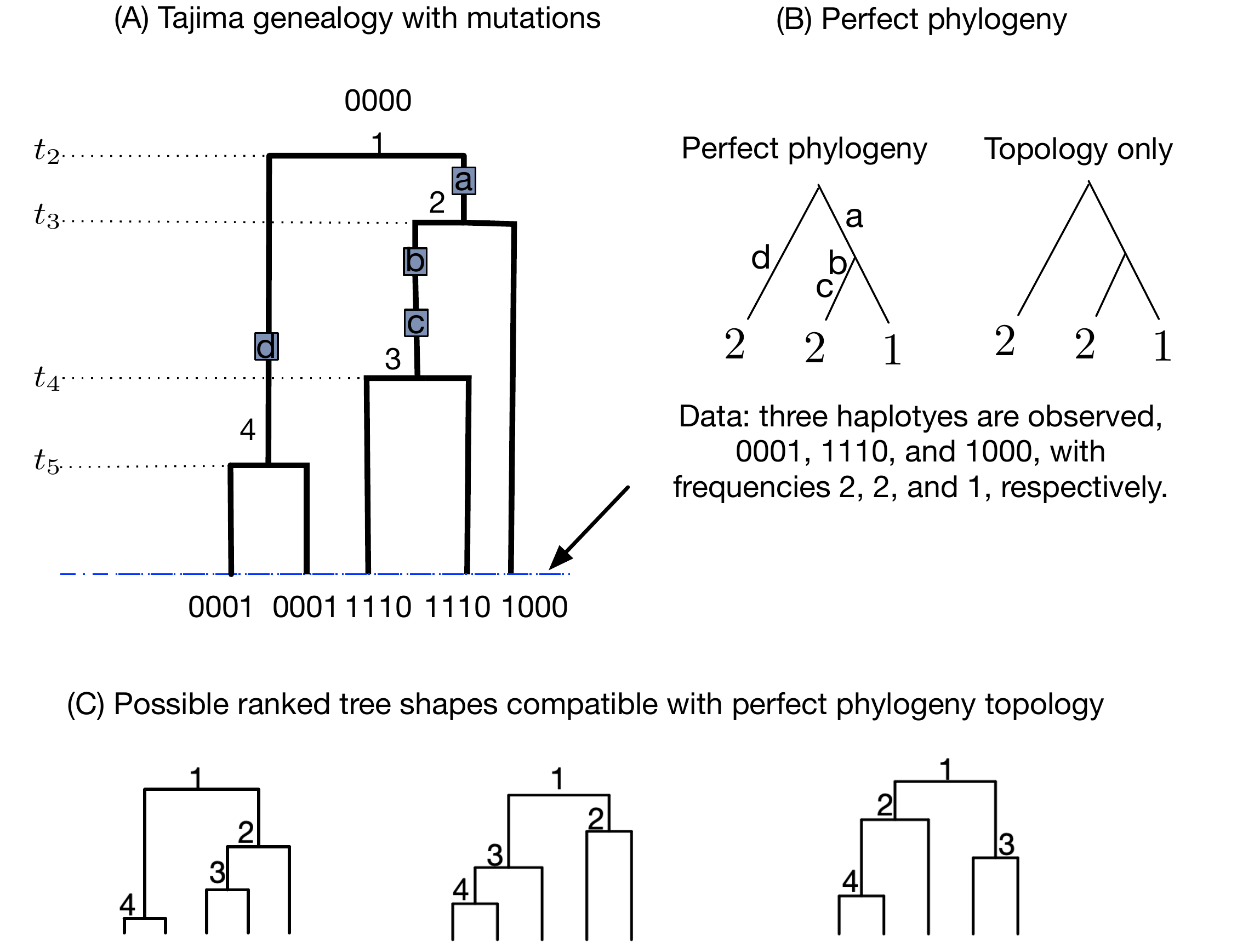}
%  % \includegraphics[width=1.0\textwidth]{Figures/tree_s.pdf}
  \end{center}
\caption{\small{\textbf{Tajima coalescent and infinitely-many-sites generative model of binary molecular data.} \textbf{(A)} A Tajima genealogy of 5 individuals, with 4 superimposed mutations depicted as gray squares. The root is labeled by the ancestral type $0000$, and the leaves are labeled by the genetic type at each of three mutated sites. The first two leaves from left to right are labeled $0001$ because one mutation occurs in their path to the root.  The third and fourth individuals have three mutations in their path to the root and are labeled 1110; the last individual is labeled $1000$ because only one mutation occurs along its path to the root. The order and label of the mutations is unimportant; however, it is assumed that the same position, or site, in a sequence of 0s and 1s corresponds across individuals. For ease of exposition, we label the mutations a, b, c and d. The first site corresponds to mutation a, the second to b, the third to c, and the fourth to d. \textbf{(B)} Left, a perfect phylogeny representation of the observed data at the tips of (A). Data consist of 3 unique haplotypes $0001$, $1110$ and $1000$, with frequencies 2, 2, and 1. The corresponding frequencies are the labels of tips of the perfect phylogeny. Right, perfect phylogeny topology obtained by removing the edge labels of the perfect phylogeny. \textbf{(C)} The only three ranked tree shapes compatible with the perfect phylogeny topology in (B).}}
\label{fig:data0}
\end{figure}
%%%%%%%%%%%%%%%%%%%%%%%%%%%%%%%%%%%%%%%%%%%%%%%%%%%%%%%%%%%%%

The \textbf{perfect phylogeny algorithm}, proposed by \citet{Gusfield1991}, generates a graphical representation of binary molecular sequence data that have been produced according to the infinitely-many-sites mutation model. Label individual sequences $1,2, \ldots, n$, and label mutated or ``segregating'' sites $a,b,\ldots$. The original algorithm generates a tree structure known as a \textbf{perfect phylogeny}, with tips labeled $1,2,\ldots,n$ and with edges labeled $a,b,\ldots$, that is in bijection with the observed ``labeled data.'' An edge can have no labels, one label, or more than one label. Perfect phylogenies have been central to coalescent-based inference algorithms, in which maximum likelihood or Bayesian estimation of evolutionary parameters that have given rise to the particular distribution of mutations and clade sizes on the perfect phylogeny are sought by importance sampling or Markov chain Monte Carlo \citep{griffiths_sampling_1994,StephensDonnelly2000,Palacios2019,cappello2020tajima}.

In this study, we assume that individual sequences are not uniquely labeled, but instead, are identified by their sequences of 0s and 1s, or \textbf{haplotypes}. Hence, the number of tips in our perfect phylogeny is the number of unique haplotypes, and the labels at the tips correspond to the observed frequencies of the haplotypes. For the genealogy in Figure \ref{fig:data0}A, Figure \ref{fig:data0}B shows the perfect phylogeny of the data observed at its tips. 

The key assumption of the bijection between sequence data sets and perfect phylogenies is that if a site mutates once, then all descendants of the lineage on which the mutation occurred must also have the mutation---and no other individuals will have the mutation. That is, every unique mutation, or site, partitions the sample of haplotypes into two groups: those with the mutation and those without the mutation. Hence, we group sites that induce the same partition on the haplotypes, and we call each such group of sites a \textbf{mutation group}. 

In this study, we are not concerned with the mutation labels, and hence, we remove the edge labels of the perfect phylogeny (right side of Figure \ref{fig:data0}B), so that we consider only the topology of the perfect phylogeny. In dropping the edge labels, we treat a perfect phylogeny topology as a perfect phylogeny. Henceforth, a \textbf{perfect phylogeny} is a multifurcating rooted tree with $k$ leaves, representing $k$ distinct haplotypes, each labeled by a positive integer $(n_{i})_{1\leq i \leq k}$, with $\sum^{k}_{i=1}n_{i}=n$. %We can have nodes of outdegree $\geq 0$. 
We use the symbol $\Pi_{n}$ to denote the space of perfect phylogenies of size $n$ sequences, and we use $\pi \in \Pi_{n}$ to denote a perfect phylogeny with $n$ sequences.

A perfect phylogeny $\pi$ is completely specified in a parenthetical notation, in which every leaf is represented by its label, every binary internal node is represented by $(\cdot,\cdot)$ and every multifurcating internal node is represented by $(\cdot,\ldots,\cdot)$. For example, the perfect phylogeny $\pi_{1}$ on the right in Figure \ref{fig:data0}B in parenthetical notation is $((2,1),2)$, indicating that there are two internal nodes, one merging leaves $(2,1)$ and one merging $(2,1)$ with $2$. 
%The perfect phylogeny $\pi_{3}$ on the right in 
%Figure \ref{fig:refinement} in parenthetical notation is $((4,2),(3,3))$. 

The most extreme unresolved perfect phylogeny with $n$ tips---the perfect phylogeny that is compatible with all ranked tree shapes with $n$ tips---has two representations. It can be written as a star, in which the root has degree $n$ and is the only internal node, that is, $\pi=(1,1,\ldots,1)$. It can also be written as a single node $\pi=(n)$. For our purposes, with mutations discarded, the star and single-node perfect phylogenies are indistinguishable, and they will be represented as a single-node perfect phylogeny. Details of the algorithm for generating the perfect phylogeny from binary molecular data can be found in \citet{Cappello2019}, which presents a slight modification to Gusfield's algorithm \citep{Gusfield1991}.

We say that a binary tree $T$ is \textbf{compatible} with a perfect phylogeny $\pi$ if the tree can be reduced to $\pi$ by collapsing internal edges of $T$. The number of tree shapes, ranked or unranked, that are compatible with a perfect phylogeny gives the cardinality of the corresponding posterior sampling tree space in statistical inference from sequence data sets. Given a perfect phylogeny $\pi \in \Pi_{n}$, we are interested in calculating the number of compatible ranked tree shapes with $n$ leaves and the number of compatible unranked tree shapes with $n$ leaves.

%To obtain a perfect phylogeny from binary molecular sequence data at a completely linked loci, we first group our data into $k$ unique "haplotypes" and let $n_{i}$ denote the frequency of haplotype $i$, $n=\sum^{k}_{i=1}n_{i}$. We will also group equal columns (sites) in our data to form $s$ mutation groups. In the perfect phylogeny topology:
%\begin{enumerate}
%\item Each haplotype frequency $n_{1},\ldots,n_{k}$ labels exactly one leaf of $\Pi(n_{1},\ldots,n_{k})$.
%\item Each mutation group corresponds to one internal node in $\Pi(n_{1},\ldots,n_{k})$. 
%\end{enumerate}
%
%\paragraph{}
%\begin{figure}
%  \begin{center}
%\includegraphics[scale=1.0]{Figures/string.eps}
%%  % \includegraphics[width=1.0\textwidth]{Figures/tree_s.pdf}
%  \end{center}
%\caption{\small{\textbf{A.} Perfect phylogeny toplogy. \textbf{B.} Perfect phylogeny topology with string notation}}
%\label{fig:data}
%\end{figure}

%The modified perfect phylogeny topology algorithm is described (somewhere).

\subsection{Known enumerative results}
\label{sec:known}

In advance of our effort to count tree shapes compatible with a perfect phylogeny, we state some known enumerative results for the unconstrained spaces of ranked labeled tree shapes, unranked labeled tree shapes, ranked unlabeled tree shapes, and unranked unlabeled tree shapes \citep{steel16}. 

Let $L_{n}=|\mathcal{T}^{L}_{n}|$ denote the cardinality of the space of ranked labeled trees with $n$ leaves. Then 
\begin{equation}
\label{eq:Ln}
L_{n}=\prod^{n}_{i=2}\binom{i}{2}=\frac{n!(n-1)!}{2^{n-1}}.
\end{equation}
The product is obtained by noting that for each decreasing $i$ from $n$ to $2$, there are $\binom{i}{2}$ ways of merging two labeled branches. The sequence of values of $L_n$ begins 1, 1, 3, 18, 180, 2700, 56700.

Let $X_n=|\mathcal{T}^{X}_{n}|$ denote the number of unranked labeled trees with $n$ leaves. We have
\begin{equation}
\label{eq:Xn}
X_{n}=(2n-3)!! = \frac{(2n-2)!}{2^{n-1} (n-1)!}.
\end{equation}
To generate trees in $\mathcal{T}^{X}_{n}$ from trees in $\mathcal{T}^{X}_{n-1}$, a pendant edge connected to the $n$th label can be placed along each of the $2n-3$ edges of a tree with $n-1$ leaves, including an edge above the root. $X_n$ is obtained as the solution to the recursion $X_n = (2n-3)X_{n-1}$, with $X_1=1$. The sequence of values of $X_n$ begins $1, 1, 3, 15, 105, 945, 10395$.

The number of ranked tree shapes with $n$ tips is the $(n-1)$-th Euler zigzag number \citep{stanley2011enumerative}. Let  $R_{n}=|\mathcal{T}^{R}_{n}|$ denote the number of ranked tree shapes with $n$ leaves. We have the following recursion:
\begin{align}
R_1 &= 1, \, R_2= 1,  \nonumber \\
R_{n+1} &= \frac{1}{2} \sum_{k=0}^{n-1} {n-1 \choose k} R_{k+1} R_{n-k}, \, n \geq 2.
\label{eq:res1}
\end{align}
The sequence of values of $R_n$ begins 1, 1, 1, 2, 5, 16, 61. For $n \geq 1$, if the tree has $n+1$ tips, and hence $n$ interior nodes, then the root divides the tree into two ranked subtrees $T^{R}_1$ and $T^{R}_2$, where $T^{R}_1$ has $k$ interior nodes, $0 \leq k \leq n-1$, and $T^{R}_2$ has $n-1 - k$ interior nodes. There are ${n-1 \choose k}$ ways of interleaving the $k$ and $n-1-k$ interior nodes of $T^{R}_1$ and $T^{R}_2$, such that the relative orderings of the interior nodes of $T^{R}_1$ and $T^{R}_2$ are preserved in the interleaving. The number of possible ranked tree shapes with such a configuration is ${n-1 \choose k} R_{k+1} R_{n-k}$. Summing over the possibilities for $k$ from $0$ to $n-1$, and acknowledging that the identity of $T^{R}_1$ and $T^{R}_2$ can be interchanged, we get eq.~\ref{eq:res1}. 

Let $S_{n}=|\mathcal{T}_{n}|$ denote the number of unranked tree shapes with $n$ leaves. We have the following recursion:
\begin{align}
S_1 &= 1, \nonumber \\
S_{2n-1} &= \sum^{n-1}_{k=1}S_{k}S_{2n-1-k}, \, n \geq 2, \\
S_{2n} &= \bigg(\sum^{n-1}_{k=1}S_{k}S_{2n-k}\bigg)+\frac{1}{2}S_{n}(S_{n}+1), \, n \geq 1.
\end{align}
$S_{n}$ is the $n$th Wedderburn-Etherington number \citep{harding71}. The sequence begins 1, 1, 1, 2, 3, 6, 11. When the number of leaves is $2n-1$, the root divides the tree shape into two subtree shapes $T_{1}$ and $T_{2}$ with $k$ and $2n-1-k$ leaves, for $k=1,2,\ldots,n-1$. When the number of leaves is even, the root divides the tree shape into subtree shapes with $k$ and $2n-k$ leaves for $k=1,2,\ldots,n-1$ or two subtree shapes with $n$ leaves; these tree shapes are indistinguishable in $S_{n}$ cases and distinguishable in $\frac{1}{2}S_n(S_{n}-1)$ cases.
 
\section{Enumeration for binary perfect phylogenies}

To count ranked and unranked tree shapes compatible with a perfect phylogeny, we first consider binary perfect phylogenies: those perfect phylogenies for which the outdegree of any node, traversing from root to tips, is either 0 (leaves or taxa) or 2 (internal nodes). We then consider multifurcating perfect phylogenies in Section \ref{sec:four}.

%{\bf XXX Julia, I feel there may be something wrong with the lattice part, but now I am confused and it could be just my fault. I would suggest to leave as it is. Let's see what the referees will say about that. XXX} \textcolor{red}{Maybe we can figure it out together? Can we zoom? just tell me when this weekend or next week works? 8am california time is good for me}

\subsection{Lattice structure of binary perfect phylogenies}
%\setcounter{section}{1}
%{\bf XXX I WOULD CHANGE THE TITLE, SOMETHING LIKE "LATTICE STRUCTURE AND ENUMERATION OF BINARY PERFECT PHILOGENIES" SHOULD WORK, AND SPLIT THE SECTION INTO TWO PARTS: "LATTICE OF BINARY PERFECT PHYLOGENIES" AND "COMPUTING THE NUMBER OF TREES COMPATIBLE WITH A BINARY PERFECT PHYLOGENY" XXX}

%
%\begin{defn}
%We say that a \textbf{binary perfect phylogeny topology $\beta$ is compatible with a perfect phylogeny topology $\Pi$} if $\beta$ can be reduced to $\Pi$ ($\beta \rightsquigarrow \Pi)$ by collapsing internal edges. We use the symbol $\mathcal{B}(\Pi)=\{\beta: \beta \rightsquigarrow \Pi\}$ to denote the set of binary perfect phylogenies compatible with $\Pi$.
%\end{defn}

The binary perfect phylogenies for a set of $n$ tips possess a structure that will assist in enumerating binary ranked and unranked trees compatible with a set of sequences. In particular, we can make the set $\Pi_{n}$ of all binary perfect phylogenies of $[n]$ into a \textbf{poset} by defining $\pi \leq \sigma$ if either $\sigma$ is the same as $\pi$, or if $\sigma$ can be obtained by sequentially collapsing pairs of pendant edges, or cherries, of $\pi$. We then say $\pi$ is a \textbf{refinement} of $\sigma$. For example, $\pi=(2,3)$ refines $\sigma=(5)$. We say that two binary perfect phylogenies in $\Pi_{n}$ are \textbf{comparable} if they are equal or if one is a refinement of the other. An example of two perfect phylogenies that are not comparable is $\pi=(2,3)$ and $\sigma=(4,1)$.

Given two binary perfect phylogenies $\pi_{1}$ and $\pi_{2}$ in $\Pi_{n}$, their \textbf{meet}, denoted $\pi_{1} \wedge \pi_{2}$, is the largest perfect phylogeny that refines both $\pi_{1}$ and $\pi_{2}$. Similarly, the \textbf{join} of two binary perfect phylogenies $\pi_{1}  \vee \pi_{2}$ is the smallest perfect phylogeny that is refined by both $\pi_{1}$ and $\pi_{2}$. Formal definitions of these notions appear in Definition \ref{def:binope}.

\begin{figure}
\centering
\includegraphics[scale=0.3]{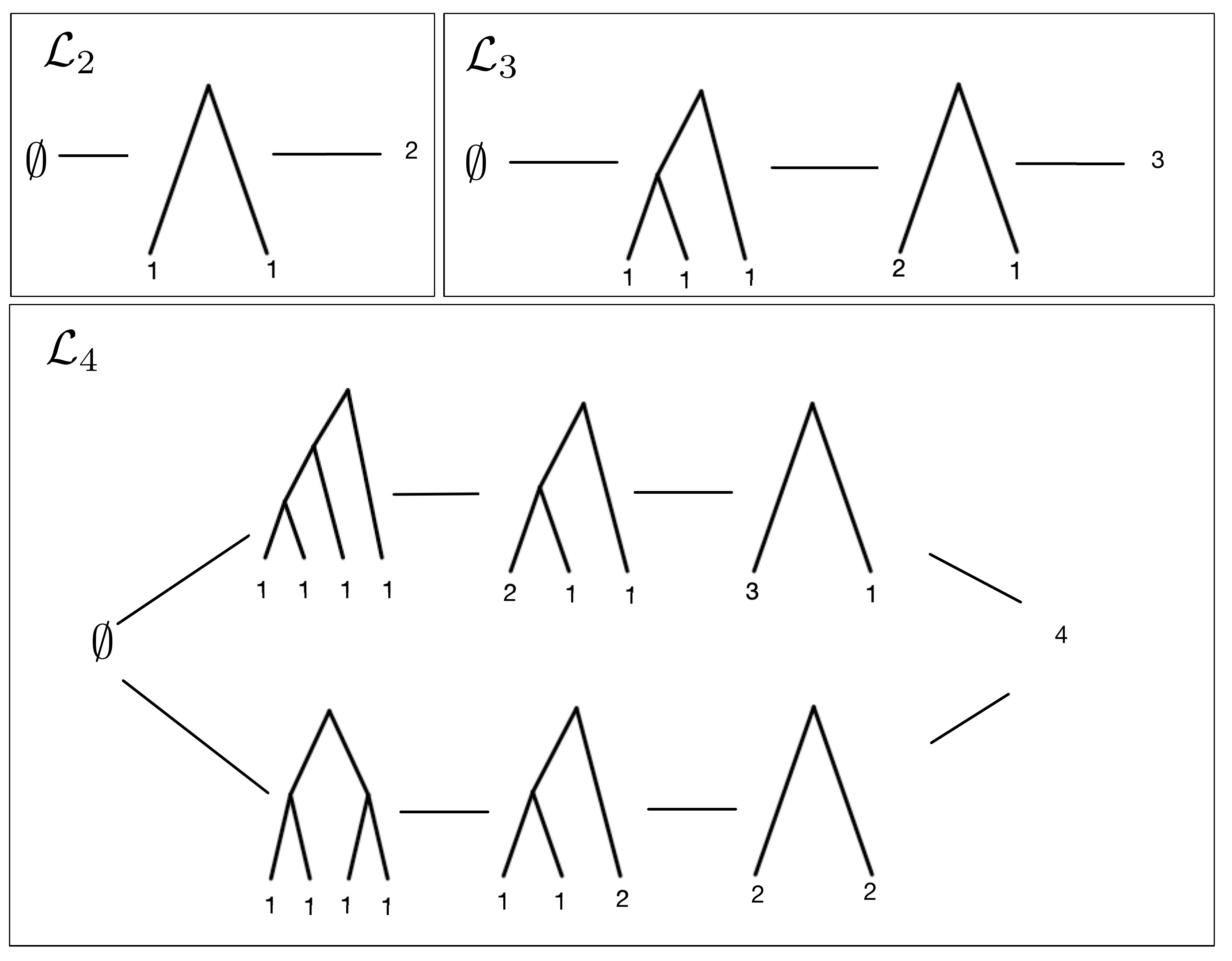}
\caption{\small{\textbf{Hasse diagrams of the lattices of binary perfect phylogenies with $n=2$, $3$, and $4$ taxa.} }}
\label{fig:hasse1}
\end{figure}

\begin{figure}
\centering
\includegraphics[scale=0.3]{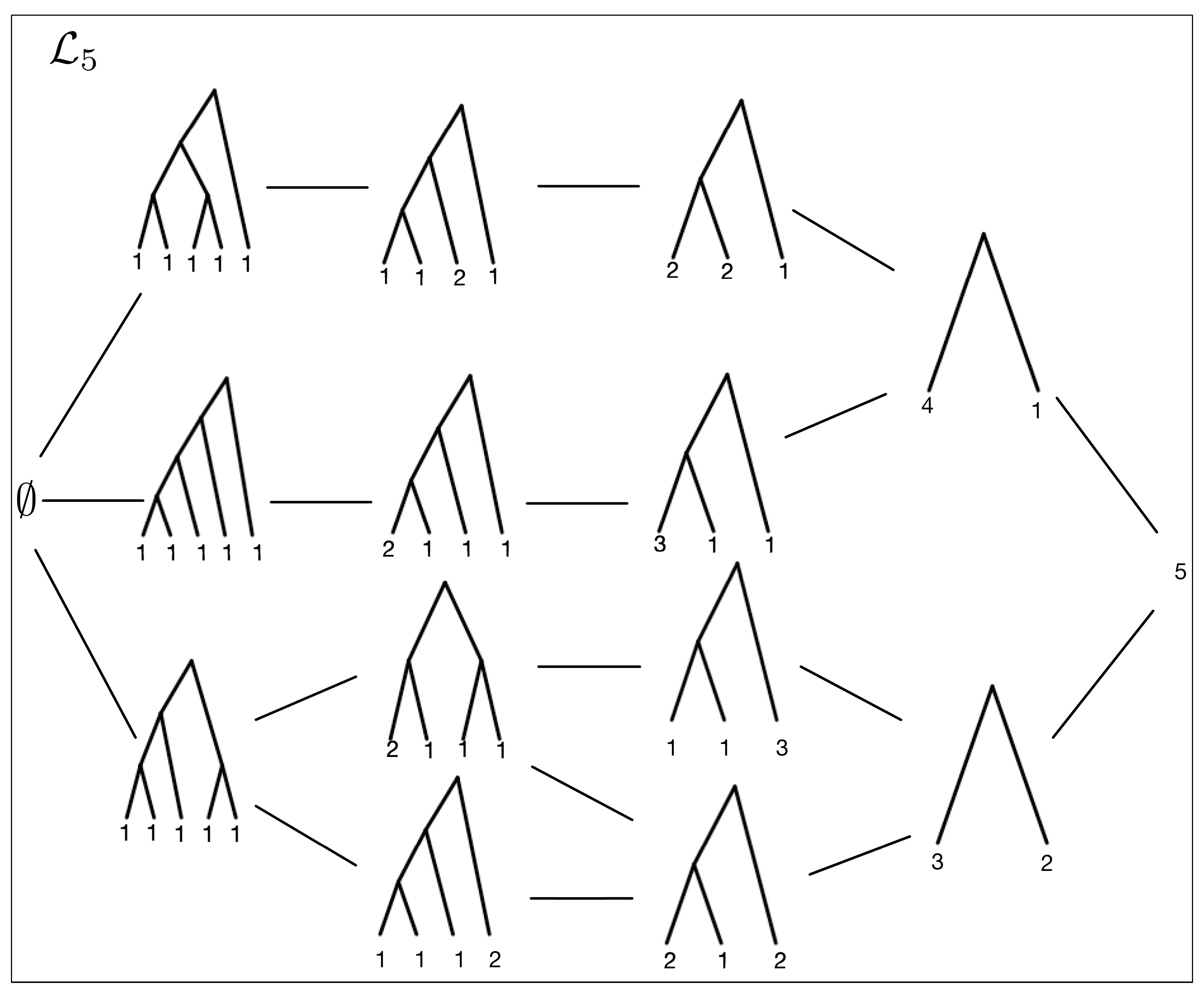}
\caption{\small{\textbf{Hasse diagram of the lattice of binary perfect phylogenies with $n=5$ taxa. }}}
\label{fig:hasse2}
\end{figure}

%The existence of these two operations is shown in Lemmas \ref{lemma_1} and \ref{lemma_2}. 
Under the meet and join operations, we will see in Theorem \ref{thm:lattice} that the poset $\Pi_{n} \cup \{\emptyset\}$ is a \textbf{lattice} $\mathcal{L}_n = (\Pi_{n} \cup \{\emptyset\}, \wedge, \vee)$. As a lattice, $\mathcal{L}_n$ possesses a \textbf{Hasse diagram} with a minimal and a maximal element. The \textbf{maximal} element of $\mathcal{L}_n$ is the single node perfect phylogeny $(n)$ and the \textbf{minimal} element is $\emptyset$. Figures \ref{fig:hasse1} and \ref{fig:hasse2} show the Hasse diagrams of $\mathcal{L}_2$, $\mathcal{L}_3$, $\mathcal{L}_4$, $\mathcal{L}_5$.

%%%%%%%%%%%%%%%%%%%%%%%%%%%%% Figure 4 %%%%%%%%%%%%%%%%%%%%%%%%%%%%%%
%%%%%%%%%%%%%%%%%%%%%%%%%%%%%%%%%%%%%%%%%%%%%%%%%%%%%%%%%%%%%%%%%%%%%

%%%%%%%%%%%%%%%%%%%%%%%%%%%% Figure 5 %%%%%%%%%%%%%%%%%%%%%%%%%%%%%%%
%%%%%%%%%%%%%%%%%%%%%%%%%%%%%%%%%%%%%%%%%%%%%%%%%%%%%%%%%%%%%%%%%%%%%

\begin{defn} \label{def:binope}
\textbf{Binary perfect phylogeny operations}. We define the binary perfect phylogeny symmetric operations $\wedge, \vee: (\cup_{n \geq 1} \Pi_{n} \cup \{ \emptyset \}) \times (\cup_{n \geq 1} \Pi_{n} \cup \{ \emptyset \}) \rightarrow (\cup_{n \geq 1} \Pi_{n} \cup \{ \emptyset \})$, where $\Pi_{n}$ is the space of binary perfect phylogenies of $n$ leaves, as follows:
%TODO AB: We only need \#1--\#3 (base cases) and \#7 (general case). 
\begin{enumerate}
\item $\pi \wedge \emptyset =\emptyset$, for all $\pi \in \Pi_{n}$.
\item $\pi \vee \emptyset = \pi$, for all $\pi \in \Pi_{n}$.
\item $\pi \wedge (n) = \pi$, for all $\pi \in \Pi_{n}$. 
\item $\pi \vee (n) = (n)$, for all $\pi \in \Pi_{n}$.
\item $\pi_1 \wedge \pi_2 = \emptyset$, for all $\pi_1 \in \Pi_{n_1}, \pi_2 \in \Pi_{n_2}$, with $n_1 \neq n_2$.
\item $\pi_1 \vee \pi_2 = \emptyset$, for all $\pi_1 \in \Pi_{n_1}, \pi_2 \in \Pi_{n_2}$, with $n_1 \neq n_2$.

%\item Let $\pi_{1}=(n_{1})$ and $\pi_{2}=(n_{2})$ two single node perfect phylogenies, then
%\begin{equation*}
%\pi_{1}\wedge \pi_{2}=(n_{1}) \wedge (n_{2})=\begin{cases} (n_{1}) & \text{ if  } n_{1}=n_{2} \\
%\emptyset & \text{ if  } n_{1} \neq n_{2}\\
%\end{cases}
%\end{equation*}
%and
%\begin{equation*}
%\pi_{1}\vee \pi_{2}=(n_{1}) \vee (n_{2})=\begin{cases} (n_{1}) & \text{ if  } %n_{1}=n_{2} \\
%\emptyset & \text{ if  } n_{1} \neq n_{2}\\
%\end{cases}
%\end{equation*}
%\item Let $\pi_{1}=(n_{1},n_{2})$ and $\pi_{2}=(n_{3})$ a cherry perfect phylogeny and a single node perfect phylogeny respectively, then
%\begin{equation*}
%\pi_{1} \wedge \pi_{2}=(n_{1},n_{2}) \wedge (n_{3})=\begin{cases} (n_{1},n_{2}) & \text{ if  } n_{3}=n_{1}+n_{2} \\
%\emptyset & \text{ otherwise } \\
%\end{cases}
%\end{equation*}
%and
%\begin{equation*}
%\pi_{1} \vee \pi_{2}=(n_{1},n_{2}) \vee (n_{3})=\begin{cases} (n_{3}) & \text{ if  } n_{3}=n_{1}+n_{2} \\
%\emptyset & \text{ otherwise } \\
%\end{cases}
%\end{equation*}
\item Let $\pi_{1}=(n_{1},n_{2})$ and $\pi_{2}=(n_{3},n_{4})$ be two perfect phylogenies in $\Pi_{n}$ with $n_{1}+n_{2}=n_{3}+n_{4}=n$. Then
%\begin{equation*}
%\pi_{1} \wedge \pi_{2}=(n_{1},n_{2}) \wedge (n_{3},n_{4})=\begin{cases} (n_{1},n_{2}) & \text{ if  } n_{1}=n_{3} \text{ or } n_{1}=n_{4} \\
%\emptyset & \text{ otherwise } \\
%\end{cases}
%\end{equation*}
%and
\begin{equation*}
\pi_{1} \vee \pi_{2}=(n_{1},n_{2}) \vee (n_{3},n_{4})=\begin{cases} (n_{1},n_{2}) & \text{ if  } n_{1}=n_{3} \text{ or } n_{1}=n_{4}\\
(n) & \text{ otherwise}. \\
\end{cases}
\end{equation*}
\item For all $\pi_{1}$, $\pi_{2}$, $\pi_{3}$, $\pi_{4}$ with $(\pi_{1},\pi_{2}) \in \Pi_{n}$ and $(\pi_{3},\pi_{4}) \in \Pi_{n}$,
\begin{equation*}
(\pi_{1},\pi_{2}) \wedge (\pi_{3},\pi_{4})=
(\pi_{1}\wedge \pi_{3}, \pi_{2} \wedge \pi_{4}) \vee
 (\pi_{1}\wedge \pi_{4}, \pi_{2} \wedge \pi_{3}),
\end{equation*}
with the convention that $(\pi,\emptyset)=\emptyset$. That is, the meet of two perfect phylogenies is the join of the two perfect phylogenies formed by merging two subtrees at the root. These four subtrees (two per newly formed perfect phylogeny) correspond to the meets of all pairs of subtrees, one from each of the original perfect phylogenies. 

\item For all $\pi_{1}$, $\pi_{2}$, $\pi_{3}$, $\pi_{4}$ with $(\pi_{1},\pi_{2}) \in \Pi_{n}$ and $(\pi_{3},\pi_{4}) \in \Pi_{n}$, $\pi_{i}\in \Pi_{n_{i}}$ for $i=1,2,3,4$.%\in \Pi_{n}$
\begin{eqnarray}
(\pi_{1},\pi_{2}) \vee (\pi_{3},\pi_{4})=\begin{cases}
(n) & \text{ if } n_{1}\neq n_{3} \text{ and } n_{1} \neq n_{4}\\
(\pi_{1}, \pi_{2} \vee \pi_{4}) & \text{ if } \pi_{1}=\pi_{3}\\
(\pi_{1}, \pi_{2} \vee \pi_{3}) & \text{ if } \pi_{1}=\pi_{4}\\
(\pi_{2}, \pi_{1} \vee \pi_{4}) & \text{ if } \pi_{2}=\pi_{3}\\
(\pi_{2}, \pi_{1} \vee \pi_{3}) & \text{ if } \pi_{2}=\pi_{4}\\
(\pi_{1}\vee \pi_{3}, \pi_{2} \vee \pi_{4}) \wedge
 (\pi_{1}\vee \pi_{4}, \pi_{2} \vee \pi_{3}) & \text{ otherwise},
\end{cases} \nonumber
\end{eqnarray}
with the convention that $(\pi,\emptyset)=\emptyset$. That is, the join of two perfect phylogenies is the meet of the two perfect phylogenies formed by merging two subtrees at the root. These four subtrees (two per newly formed perfect phylogeny) correspond to the joins of all pairs of subtrees, one from each of the original perfect phylogenies. In the particular case that the two original perfect phylogenies share one of the subtrees descending from the root, then the join of the two perfect phylogenies is the perfect phylogeny that merges, at the root, the shared subtree with the join of the two different subtrees, one from each of the original perfect phylogenies. In the case that no two pairs of subtrees, one from each of the original perfect phylogenies, have the same size, the join is the maximal single node perfect phylogeny $(n)$. 
%\item For any $\pi_{1},\pi_{3} \in \Pi_{n_{1}}$ and $\pi_{2}, \pi_{4} \in \Pi_{n_{2}}$ with $n_{1} \ne n_{2}$
%\begin{equation*}
%(\pi_{1},\pi_{2}) \wedge (\pi_{3},\pi_{4})=
%(\pi_{1}\wedge \pi_{3}, \pi_{2} \wedge \pi_{4})  
%\end{equation*}
%and
%\begin{equation*}
%(\pi_{1},\pi_{2}) \vee (\pi_{3},\pi_{4})=
%(\pi_{1}\vee \pi_{3}, \pi_{2} \vee \pi_{4}) 
%\end{equation*}

%with the convention that $(\pi,\emptyset)=\emptyset$
%\item For any $\pi_{1} \in \Pi_{n_{1}}$, $\pi_{2} \in \Pi_{n_{2}}$, $\pi_{3} \in \Pi_{n_{3}}$, and $\pi_{4} \in \Pi_{n_{4}}$ with $n_{1} \ne n_{2} \neq n_{3} \neq n_{4}$ and $n = n_{1} + n_{2} = n_{3} + n_{4}$,
%\begin{equation*}
%(\pi_{1},\pi_{2}) \wedge (\pi_{3},\pi_{4})=\emptyset
%\end{equation*}
%and
%\begin{equation*}
%(\pi_{1},\pi_{2}) \vee (\pi_{3},\pi_{4})=(n)=(n_{1}+n_{2})=(n_{3}+n_{4})  
%\end{equation*}
%\begin{equation}
%(\pi_{1},\pi_{2}) \wedge (\pi_{3},\pi_{4})=\begin{cases}
%(\pi_{1}\wedge \pi_{3}, \pi_{2} \wedge \pi_{4}) & \text{ if }   \pi_{1}\wedge \pi_{3} \neq \emptyset \text{ and } \pi_{2} \wedge \pi_{4} \neq \emptyset\\
% (\pi_{1}\wedge \pi_{4}, \pi_{2} \wedge \pi_{3}) &  \text{ if }   \pi_{1}\wedge \pi_{4} \neq \emptyset \text{ and } \pi_{2} \wedge \pi_{3} \neq \emptyset\\
% \emptyset & \text{otherwise}
% \end{cases}
%\end{equation}
\item For all $\pi_{1},\pi_{2}, \pi_{3} \in \Pi_{n},$
\begin{equation*}
    \pi_{1} \wedge (\pi_{2} \vee \pi_{3})=(\pi_{1} \wedge \pi_{2}) \vee (\pi_{1} \wedge \pi_{3}),
    \end{equation*}
and
\begin{equation*}
    \pi_{1} \vee (\pi_{2} \wedge \pi_{3})=(\pi_{1} \vee \pi_{2}) \wedge (\pi_{1} \vee \pi_{3}).
    \end{equation*}

%\item\textcolor{red}{ Let $\pi, \sigma \in \Pi_{n}$ be two perfect phylogenies that are not comparable, then the meet and the join operations are such that 
%\begin{equation*}\pi \wedge \sigma = \gamma, \text{ where }\quad \gamma \in (\Pi_{n}  \cup \{\emptyset\}) \setminus \{\pi,\sigma\}, \quad \pi \vee \gamma = \pi \quad \text{ and }\quad \sigma \vee \gamma =\sigma,
%\end{equation*}
%and
%\begin{equation*}\pi \vee \sigma = \rho, \text{ where }\quad \rho \in (\Pi_{n}  \cup \{\emptyset\}) \setminus \{\pi,\sigma\}, \quad \pi \wedge \rho = \pi \quad \text{ and }\quad \sigma \wedge \rho =\sigma,
%\end{equation*}}

\item Let $\pi, \sigma \in \Pi_{n}$ be two perfect phylogenies that are not comparable. There exist unique $\gamma,\rho \in (\Pi_{n}  \cup \{\emptyset\}) \setminus \{\pi,\sigma\} $ such that 
\begin{equation*}\pi \wedge \sigma = \gamma, \quad \pi \vee \gamma = \pi, \quad \text{ and }\quad \sigma \vee \gamma =\sigma,
\end{equation*}
and
\begin{equation*}\pi \vee \sigma = \rho, \quad \pi \wedge \rho = \pi, \quad \text{ and }\quad \sigma \wedge \rho =\sigma.
\end{equation*}

\end{enumerate}
\end{defn}

Note that the meet and join operations are symmetric and that pairs $(\pi_1,\pi_2)$ are unordered; for convenience, we have expanded expressions in parts 7 and 9 of the definition that could potentially be simplified using the symmetry.

We illustrate the operations in Definition \ref{def:binope} by considering a series of examples.

\begin{figure}
\begin{center}
\includegraphics[scale=0.18]{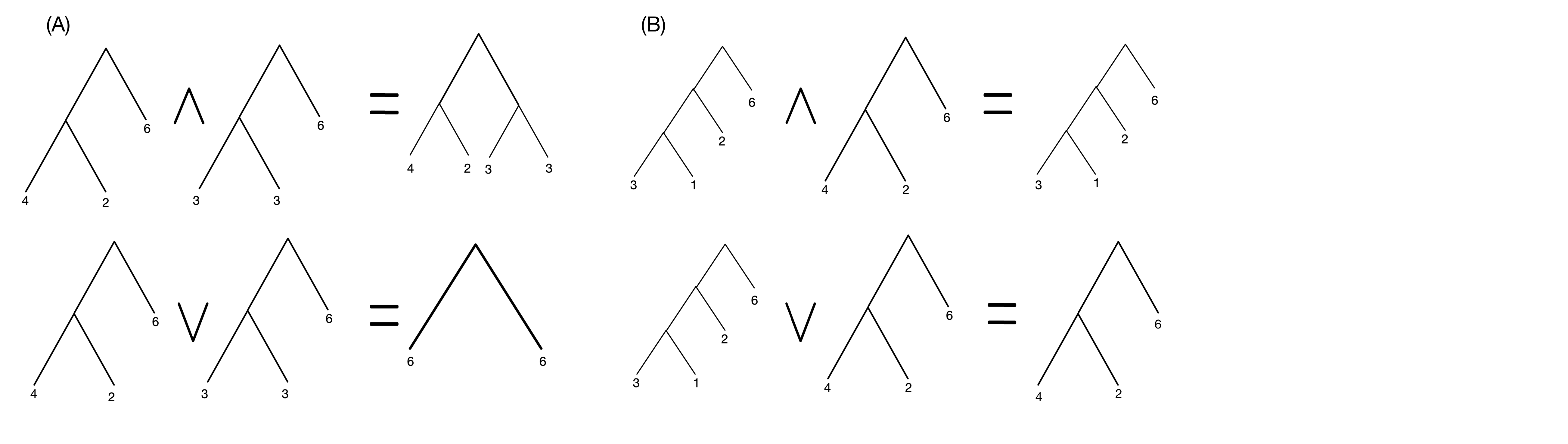}
\caption{\small{\textbf{Examples of perfect phylogeny operations.} \textbf{(A)} For perfect phylogenies $((4,2),6)$ and $((3,3),6)$, their meet is $((4,2),(3,3))$, and their join is $(6,6)$. \textbf{(B)} For perfect phylogenies $(((3,1),2),6)$ and $((4,2),6)$, their meet is $(((3,1),2),6)$ and their join is $((4,2),6)$.}}
\label{fig:refinement}
\end{center}
\end{figure}

\begin{examp}
Consider  $\pi_{1}=((4,2),6)$ and $\pi_{2}=((3,3),6)$ depicted in Figure \ref{fig:refinement}A. Their meet and join are given by:
\begin{align*}
((4,2),6) \wedge ((3,3),6)  &= ((4,2)\wedge (3,3),6\wedge 6) \vee ((4,2)\wedge 6, 6 \wedge (3,3))\text{ by Defn.~\ref{def:binope} (8)}\\
&= (\emptyset,6) \vee ((4,2),(3,3))\text{ by Defn.~\ref{def:binope} (3, 5, 8)}\\
&= \emptyset \vee ((4,2),(3,3)) \text{ by convention}\\
&=  ((4,2),(3,3)) \text{ by Defn.~\ref{def:binope} (2)}. \\[1.5ex]
%\end{align*}
%\begin{align*}
((4,2),6) \vee ((3,3),6) &= (6, (4,2)\vee (3,3)) \text{ by Defn.~\ref{def:binope} (9)}\\
&= (6,6) \text{ by Defn.~\ref{def:binope} (7).}
\end{align*}
\end{examp}

\begin{examp}
For a more complex example, consider $\pi_{1}=((3,1),2),6)$ and $\pi_{2}=((4,2),6)$ depicted in Figure \ref{fig:refinement}B.
\begin{align*}
(((3,1),2),6) \wedge ((4,2),6)  &= (((3,1),2)\wedge (4,2),6\wedge 6) \vee (((3,1),2)\wedge 6, 6 \wedge (4,2))\text{ by Defn.~\ref{def:binope} (8)}\\
& = (((3,1),2)\wedge(4,2),6) \vee (((3,1),2),(4,2)) \text{ by Defn.~\ref{def:binope} (3)} \\
& = ( ((3,1) \wedge 4, 2\wedge 2),6) \vee (((3,1),2),(4,2)) \text{ by Defn.~\ref{def:binope} (2, 5, 8)} \\ 
& = (((3,1),2),6) \vee (((3,1),2),(4,2))
\text{ by Defn.~\ref{def:binope} (3)} \\
&= (((3,1),2),6) \text{ by Defn.~\ref{def:binope} (4, 9)}. \\[1.5ex]
%\end{align*}
%\begin{align*}
((3,1),2),6) \vee ((4,2),6) &= (((3,1),2)\vee (4,2),6) \text{ by Defn.~\ref{def:binope} (9)}  \\
&= ((4,2),6) \text{ by Defn.~\ref{def:binope} (4, 9)}.
\end{align*}
\end{examp}

%%%%%%%%%%%%%%%%%%%%%%%%%%%% Figure 6 %%%%%%%%%%%%%%%%%%%%%%%%%%%%%% 

To make use of the operations $\wedge$ and $\vee$ for counting binary ranked and unranked trees compatible with a perfect phylogeny, we need a theorem that shows that the two operations $\wedge$ and $\vee$ induce the same order. That is, we will show that $(\Pi_{n} \cup \{\emptyset\}, \wedge,\vee)$ is a lattice. 

A \textit{lattice} \citep{nation1998notes} is an algebra $\mathcal{L}(L,\wedge,\vee)$ satisfying, for all $x,y,z \in L$,
\begin{enumerate}
\item $x \wedge x =x$ and $x \vee x=x$,
\item $x \wedge y =y \wedge x$ and $x \vee y=y \vee x$,
\item $x \wedge (y \wedge z) = (x \wedge y) \wedge z$ and $x \vee (y \vee z)=(x\vee y) \vee z$,
\item $x \wedge (x \vee y)=x$ and $ x\vee (x \wedge y)=x$.
\end{enumerate}
In the Appendix, we verify these conditions for $(\Pi_{n} \cup \{\emptyset\},\wedge,\vee)$, giving the following theorem.
\begin{theorem}\label{thm:lattice}
$(\Pi_{n} \cup \{\emptyset\},\wedge,\vee)$ is a lattice.
\end{theorem}

%(2) $\pi \wedge \sigma=\sigma \wedge \pi$ and $\pi \vee \sigma=\sigma \vee \pi$, (3) $\pi \wedge (\sigma \wedge \phi)=(\pi \wedge \sigma) \wedge \phi$ and $\pi \vee (\sigma \vee \phi)=(\pi \vee \sigma) \vee \phi$ and  $\pi \wedge (\pi \vee \sigma)=\pi$ and $\pi \vee (\pi \wedge \sigma)=\pi$ . Which I think are obvious. Ok, so we can use the meet and join operations for counting. Do you know what else can we do exploiting the fact that is a lattice? Some inversion formula that can be useful?
%
\subsection{Unranked unlabeled tree shapes compatible with a binary perfect phylogeny} 
\label{sec:unranked}

%{\bf XXX HERE THE NEW SECTION: "COMPUTING  THE  NUMBER  OF  TREES COMPATIBLE WITH A BINARY PERFECT PHYLOGENY" XXX} 
 
With the lattice structure of the binary perfect phylogenies established, we are now equipped to calculate the number of compatible unranked unlabeled tree shapes with $n$ leaves. Notice that an unranked unlabeled tree shape can be transformed into a perfect phylogeny with the same number of tips by assigning the count 1 to all leaves. We use $\mathcal{P}(T_{n})$ to denote the perfect phylogeny with $n$ tips that corresponds to the unranked unlabeled tree shape $T_{n}$. 

\begin{defn}
\label{def:treeshape_comp}
\textbf{Unranked unlabeled tree shape $T_{n}$ compatible with a perfect phylogeny $\pi \in \Pi_{n}$.} An unranked unlabeled tree shape with $n$ leaves, $T_{n}$, is compatible with a perfect phylogeny $\pi \in \Pi_{n}$, if (1) a one-to-one correspondence exists between the $k$ leaves of $\pi$ with leaf counts $n_{1},n_{2},\ldots,n_{k}$ and $k$ disjoint subtrees of $T_{n}$ containing $n_{1},n_{2},\ldots,n_{k}$ leaves, respectively; and (2) $\mathcal{P}(T_{n})\leq \pi$, that is, $\mathcal{P}(T_{n})$ is a refinement of $\pi$. 
\end{defn}

\begin{figure}
\centering
\includegraphics[scale=0.60]{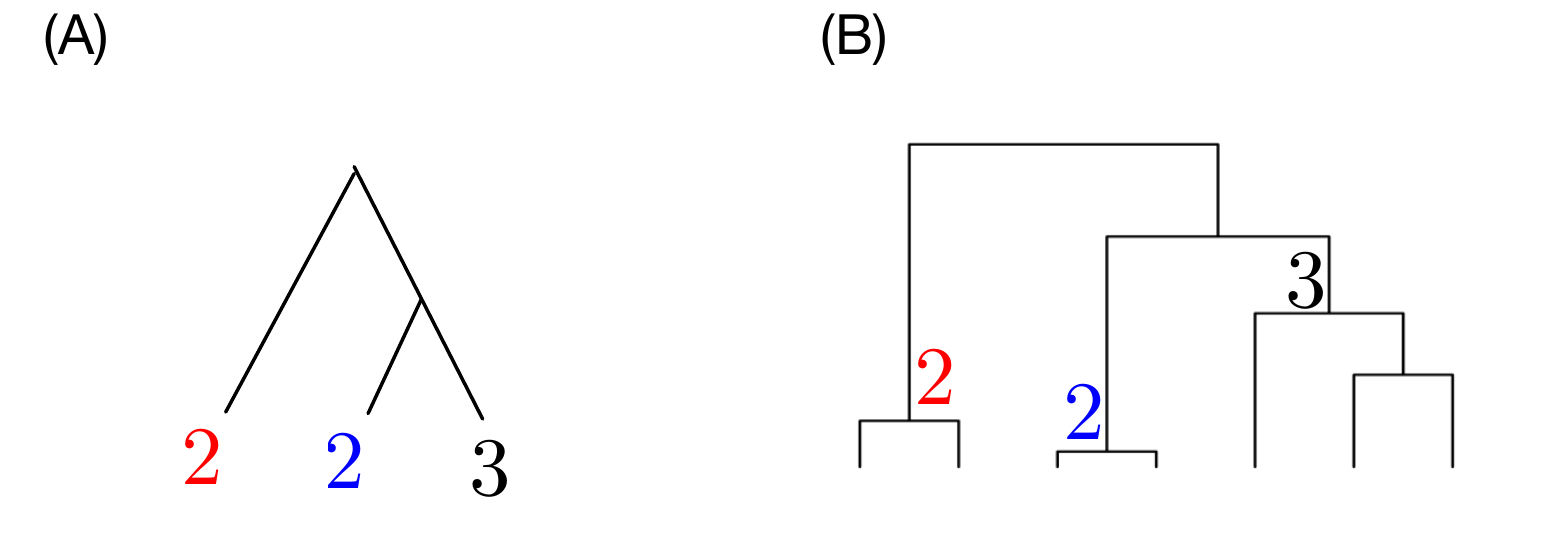}
\caption{\small{\textbf{Example of a tree shape compatible with a perfect phylogeny.} \textbf{(A)} A perfect phylogeny. \textbf{(B)} An unranked unlabeled tree shape that is compatible with the perfect phylogeny in (A). The numbers indicate the one-to-one correspondence described in Definition \ref{def:treeshape_comp}.}}.
\label{fig:compat}
\end{figure}

We use the symbol $\mathcal{G}_{c}(\pi)=\{T_{n}:T_{n} \rightsquigarrow \pi\}$ to denote the set of unranked unlabeled tree shapes compatible with a perfect phylogeny $\pi \in \Pi_{n}$. For a perfect phylogeny $\pi$ consisting of a single leaf with leaf count $n$, the number of compatible unranked unlabeled tree shapes is simply the number of unranked unlabeled tre shapes of size $n$, or $|\mathcal{G}_{c}(\pi)| = S_n$. Figure \ref{fig:compat} shows an example of an unranked unlabeled tree shape compatible with a perfect phylogeny of sample size 7.

%%%%%%%%%%%%%%%%%%%%%%%%%%%%%%%%%%%%%%%% Figure 7 %%%%%%%%%%%%%%%%%%
%%%%%%%%%%%%%%%%%%%%%%%%%%%%%%%%%%%%%%%%%%%%%%%%%%%%%%%%%%%%%%%%%%%%

\begin{prop}
For $n_1,n_2 \geq 1$, the number of unranked unlabeled tree shapes compatible with a cherry perfect phylogeny $(n_{1},n_{2}) \in \Pi_{n}$ is
\begin{eqnarray}
|\mathcal{G}_{c}((n_{1},n_{2}))|=\begin{cases}
S_{n_{1}}S_{n_{2}} &\text{if } n_{1}  \neq n_{2}\\
\frac{1}{2}S_{n_{1}}(S_{n_{1}}+1) & \text{if } n_{1} = n_{2}.
\end{cases}
\end{eqnarray}
\label{prop6}
\end{prop}
\begin{proof}
By Definition \ref{def:treeshape_comp}, an unranked unlabeled tree shape is compatible with the perfect phylogeny $\pi = (n_1,n_2)$ if it possesses two subtrees, one with $n_1$ leaf descendants and another with $n_2$ leaf descendants. Decomposing an unranked unlabeled tree shape at its root, the number of shapes with this property is $S_{n_{1}}S_{n_{2}}$ for $n_1 \neq n_2$ and $\frac{1}{2}S_{n_{1}}(S_{n_{1}}+1)$ for $n_1=n_2$.
\end{proof}

\begin{prop}
For $n_1,n_2 \geq 1$ and $\pi_{1} \in \Pi_{n_{1}}$, $\pi_{2} \in \Pi_{n_{2}}$, the number of unranked unlabeled tree shapes compatible with a binary perfect phylogeny $\pi=(\pi_{1},\pi_{2}) \in \Pi_{n}$  is
\begin{eqnarray}
|\mathcal{G}_{c}((\pi_{1},\pi_{2}))|= \begin{cases}
|\mathcal{G}_{c}(\pi_{1})| \, |\mathcal{G}_{c}(\pi_{2})|-\frac{1}{2}|\mathcal{G}_{c}(\pi_{1} \wedge \pi_{2})| \, (|\mathcal{G}_{c}(\pi_{1} \wedge \pi_{2})|-1) & \text{if } \pi_{1} \wedge \pi_{2} \neq \emptyset\\
|\mathcal{G}_{c}(\pi_{1})| \, |\mathcal{G}_{c}(\pi_{2})| & \text{if } \pi_{1} \wedge \pi_{2}= \emptyset. \\
\end{cases}
\end{eqnarray}
\label{prop7}
\end{prop}
\begin{proof}
If $\pi_{1} \wedge \pi_{2}=\emptyset$, then no tree shapes are compatible with both $\pi_{1}$ and $\pi_{2}$. Hence, the number of tree shapes compatible with $(\pi_{1},\pi_{2})$ is simply the product of the number of tree shapes compatible with $\pi_{1}$ and the number of tree shapes compatible with $\pi_{2}$. 

If $\pi_{1} \wedge \pi_{2} \neq \emptyset$, then certain tree shapes can be compatible with both $\pi_{1}$ and $\pi_{2}$, i.e., compatible with $\pi_{1} \wedge \pi_{2}$. We sum four quantities. (1) Consider the set of tree shapes compatible with both perfect phylogenies $\pi_{1}$ and $\pi_{2}$. They can either be assigned the same tree shape, in $|\mathcal{G}_{c}(\pi_{1} \wedge\pi_{2})|$ ways, or they can be assigned different tree shapes, in $\frac{1}{2}(|\mathcal{G}_{c}(\pi_{1} \wedge\pi_{2})|^{2}-|\mathcal{G}_{c}(\pi_{1} \wedge\pi_{2})|)$ ways, resulting in $\frac{1}{2}|\mathcal{G}_{c}(\pi_{1} \wedge\pi_{2})|(|\mathcal{G}_{c}(\pi_{1} \wedge\pi_{2})|+1)$ tree shapes. (2) If $\pi_{2}$ is a refinement of $\pi_{1}$ and $\pi_{1} \neq \pi_{2}$, then there are $|\mathcal{G}_{c}(\pi_{1} \wedge\pi_{2})| (|\mathcal{G}_{c}(\pi_{1})|-|\mathcal{G}_{c}(\pi_{1} \wedge\pi_{2})|)$ tree shapes. (3) Similarly, if $\pi_{1}$ is a refinement of $\pi_{2}$ %{\bf XXX a refinement of what? XXX} 
and $\pi_{1} \neq \pi_{2}$, then there are $|\mathcal{G}_{c}(\pi_{1} \wedge\pi_{2})| (|\mathcal{G}_{c}(\pi_{2})|-|\mathcal{G}_{c}(\pi_{1} \wedge\pi_{2})|)$. (4) If $\pi_{1}$ and $\pi_{2}$ are not comparable, that is, if neither is a refinement of the other, then there are $(|\mathcal{G}_{c}(\pi_{1})|- |\mathcal{G}_{c}(\pi_{1} \wedge\pi_{2})|) (|\mathcal{G}_{c}(\pi_{2})|-|\mathcal{G}_{c}(\pi_{1} \wedge\pi_{2})|)$ tree shapes. Scenarios (2), (3), and (4) are mutually exclusive, and only one of the quantities in (2), (3), and (4) is nonzero; summing the four quantities gives the result.
\end{proof}

Propositions \ref{prop6} and \ref{prop7} provide a recursive formula for calculating the number of tree shapes compatible with a binary perfect phylogeny. For example, examining Figure \ref{fig:refinement}A, the number of tree shapes compatible with $(4,2)$ is $S_{4}S_{2}=2$, and the number of tree shapes compatible with $((4,2),6)$ is $|\mathcal{G}_{c}(4,2)| \, |\mathcal{G}_{c}(6)|- \frac{1}{2}|\mathcal{G}_{c}(4,2)| \, (|\mathcal{G}_{c}(4,2)|-1) =(2) (6)-\frac{1}{2}(2)(1)=11.$ Table 1 shows the number of tree shapes compatible with certain perfect phylogenies of sample size $10$.

\subsection{Ranked unlabeled tree shapes compatible with a binary perfect phylogeny} 
\label{sec:ranked}

Next, for a binary perfect phylogeny, we compute the number of compatible ranked unlabeled tree shapes with $n$ leaves.

\begin{defn}
\label{def:ranked_comp}
\textbf{Ranked unlabeled tree shape $T^{R}_{n}$ compatible with a perfect phylogeny $\pi \in \Pi_{n}$}. A ranked unlabeled tree shape with $n$ leaves, $T^{R}_{n}$, is compatible with a perfect phylogeny $\pi \in \Pi_{n}$  if the unranked unlabeled tree shape $T_{n}$ obtained by removing the ranking from ${T}^{R}_{n}$ is compatible with $\pi$.
%$\Pi$ is obtained by collapsing internal edges from the ranked tree shape $g^{T}_{n}$ ($g^{T}_{n} \rightsquigarrow \Pi$). We use the symbol $\mathcal{G}^{T}_{c}(\Pi)=\{g^{T}_{n}:g^{T}_{n} \rightsquigarrow \Pi\}$ to denote the set of ranked tree shapes compatible with $\Pi$.
\end{defn}

%with leaves $n_{1},\ldots,n_{k}$, $n=\sum^{k}_{i=1}n_{i}$,

\begin{prop}
For $n_1, n_2 \geq 1$, the number of ranked unlabeled tree shapes compatible with a cherry perfect phylogeny $(n_{1},n_{2}) \in \Pi_{n}$ is
\begin{eqnarray}
|\mathcal{G}^{T}_{c}((n_{1},n_{2}))|=\begin{cases}
\binom{n_{1}+n_{2}-2}{n_{1}-1}R_{n_{1}}R_{n_{2}} &\text{if } n_{1}  \neq n_{2}\\
\frac{1}{2}\binom{2n_{1}-2}{n_{1}-1}R^{2}_{n_{1}} & \text{if } n_{1} = n_{2}.
\end{cases}
\end{eqnarray}
\label{prop9}
\end{prop}
\begin{proof}
By Definition \ref{def:ranked_comp}, a ranked unlabeled tree shape $T^R$ is compatible with the perfect phylogeny $\pi = (n_1,n_2)$ if the associated unranked unlabeled tree shape $T$ obtained by removing the ranking of $T^R$ is compatible with $\pi$. By Definition \ref{def:treeshape_comp}, the unranked unlabeled tree shape $T$ is compatible with the perfect phylogeny $\pi = (n_1,n_2)$ if it possesses two subtrees, one with $n_1$ leaf descendants and another with $n_2$ leaf descendants. 

We decompose a ranked unlabeled tree at its root into subtrees of size $n_1$ and $n_2$. If $n_{1} \neq n_{2}$, then the  $n_{1}-1$ interior nodes of the subtree with $n_{1}$ leaves and the $n_{2}-1$ interior nodes of the subtree with $n_{2}$ leaves can be interleaved in $\binom{n_{1}+n_{2}-2}{n_{1}-1}$ ways. If $n_{1}=n_{2}$, then the two ranked subtrees can be the same in $R_{n_{1}}$ ways, each with $\frac{1}{2}\binom{2n_{1}-2}{n_{1}-1}$ ways of interleaving the two ranked unlabeled subtrees; the two ranked subtrees can differ in $\frac{1}{2}(R^{2}_{n_{1}}-R_{n_{1}})$ ways, each with $\binom{2n_{1}-2}{n_{1}-1}$ ways of interleaving the subtrees.
\end{proof}

\begin{prop}
For $n_1, n_2 \geq 1$ and $\pi_1 \in \Pi_{n_1}, \pi_2 \in \Pi_{n_2}$, the number of ranked unlabeled tree shapes compatible with a binary perfect phylogeny $\pi=(\pi_{1},\pi_{2}) \in \Pi_{n}$ is
%, whose root divides the binary perfect phylogeny in two binary perfect phylogenies $\pi_{1}\in \Pi_{n_{1}}$ and $\pi_{2}\in \Pi_{n_{2}}$ is:
\begin{equation}
|\mathcal{G}^{T}_{c}((\pi_1, \pi_2))|=\begin{cases}
%|\mathcal{G}^{T}_{c}(\pi_{1},\pi_{2})|=
\binom{2 n_{1}-2}{n_{1}-1}(|\mathcal{G}^{T}_{c}(\pi_{1})| \, |\mathcal{G}^{T}_{c}(\pi_{2})| -\frac{1}{2}|\mathcal{G}^{T}_{c}(\pi_{1} \wedge \pi_{2})|^{2}) & \text{ if } \pi_{1} \wedge \pi_{2} \neq \emptyset\\
\binom{n_{1}+n_{2}-2}{n_{1}-1}|\mathcal{G}^{T}_{c}(\pi_{1})| \, |\mathcal{G}^{T}_{c}(\pi_{2})| & \text{ if } \pi_{1} \wedge \pi_{2}= \emptyset.
\end{cases}
\end{equation}
%where $n=n_{1}+n_{2}$.
\label{prop10}
\end{prop}

\begin{proof}
If $\pi_{1} \wedge \pi_{2} = \emptyset$, then the number of ranked tree shapes compatible with $(\pi_{1},\pi_{2})$ is simply the product of the number of ranked tree shapes compatible with $\pi_{1}$, the number of ranked tree shapes compatible with $\pi_{2}$, and the number of ways of interleaving their rankings. 

If $\pi_{1} \wedge \pi_{2} \neq \emptyset$, then certain ranked tree shapes can be compatible with both $\pi_{1}$ and $\pi_{2}$, i.e., compatible with $\pi_{1}\wedge \pi_{2}$. We therefore have three cases: the two perfect phylogenies are the same, one is a refinement of the other (two possible ways), or neither is a refinement of the other. The cardinalities in these cases are $\frac{1}{2}|\mathcal{G}^{T}_{c}(\pi_{1} \wedge \pi_{2})|^{2}$, $|\mathcal{G}^{T}_{c}(\pi_{1} \wedge \pi_{2})| \, (|\mathcal{G}^{T}_{c}(\pi_{2})|-|\mathcal{G}^{T}_{c}(\pi_{1} \wedge \pi_{2})|)+|\mathcal{G}^{T}_{c}(\pi_{1} \wedge \pi_{2})|(|\mathcal{G}^{T}_{c}(\pi_{1})|-|\mathcal{G}^{T}_{c}(\pi_{1} \wedge \pi_{2})|)$, and $(|\mathcal{G}^{T}_{c}(\pi_{1})|-|\mathcal{G}^{T}_{c}(\pi_{1} \wedge \pi_{2})|)(|\mathcal{G}^{T}_{c}(\pi_{2})|-|\mathcal{G}^{T}_{c}(\pi_{1} \wedge \pi_{2})|)$, respectively, all multiplied by the possible number of interleavings of the rankings $\binom{2n_{1}-2}{n_{1}-1}$. %The second case follows from proposition \ref{prop3}.
\end{proof}

%{\bf XXX I HAVE REMOVED SECTION "RESULTS" THAT ONLY CONTAINED THE FOLLOWING PROPOSITION. I THINK THE NEXT RESULT FITS BETTER HERE AFTER PROPOSITION \ref{prop4}. NOTE THAT I HAVE REPHRASED THE PROPOSITION SO THAT I HOPE IT IS NOW MORE PRECISE AND COHERENT WITH ITS PROOF. I HAVE UPDATED THE PROOF. XXX}

Propositions \ref{prop9} and \ref{prop10} provide a recursive formula for calculating the number of ranked tree shapes compatible with a binary perfect phylogeny. For Figure \ref{fig:refinement}A, the number of ranked tree shapes compatible with $(4,2)$ is $(4)(2)=8$, and the number of ranked tree shapes compatible with $((4,2),6)$ is $\binom{10}{5} (|\mathcal{G}^{T}_{c}(4,2)| \, |\mathcal{G}^{T}_{c}(6)|- \frac{1}{2}|\mathcal{G}^{T}_{c}(4,2)|^2) =\binom{10}{5}[(8)(16)-\frac{1}{2}(8)^{2}]=24,192$. 

Table 1 shows the number of ranked unlabeled tree shapes compatible with some of the perfect phylogenies of sample size $10$. We can observe that these numbers exceed corresponding numbers of unranked unlabeled tree shapes compatible with the perfect phylogenies, just as the numbers of ranked unlabeled tree shapes exceed the numbers of unranked unlabeled tree shapes (Section \ref{sec:known}). 

For the ranked unlabeled tree shapes compatible with a binary perfect phylogeny, we can examine the asymptotic growth of the number of compatible ranked unlabeled tree shapes in particular families of binary perfect phylogenies. For a fixed integer value $x \geq 1$, consider the family of binary perfect phylogenies $B_x(n)=(x,n-x)$ as $n$ increases. These are cherry phylogenies with labels $x$ and $n-x$ at their two leaves. Let $b_x(n)$ be the number of ranked unlabeled tree shapes compatible with $B_x(n)$. Among the integer sequences $b_1(n)$, $b_2(n)$, $b_3(n)$, $\ldots$, the next proposition shows that $b_2(n)$ has the fastest asymptotic growth.  In other words, as $n$ grows large, the value of $x$ for which the number of ranked unlabeled tree shapes compatible with the perfect phylogeny $B_x(n)$ is asymptotically largest is $x=2$.

\begin{prop}
\label{prop:asymptotic}
Among the integer sequences $b_1(n)$, $b_2(n)$, $b_3(n)$, $\ldots$, the sequence $b_2(n)$ has the fastest asymptotic growth. 
%when $n \rightarrow \infty$ the largest number of ranked tree shapes compatible with a binary perfect phylogeny occurs when there is a single bifurcation event dividing the samples in two groups of sizes $n-2$ and 2, or when there is an additional bifurcation event separating the two samples of the node of size 2. That is, when the binary perfect phylogeny is $(n-2,2)$ or $(n-2,(1,1))$.
\end{prop}
\begin{proof}

\noindent For a fixed integer value $x \geq 0$, let $\beta_x = (x+1,n-x+1)$ be a binary perfect phylogeny 
with two leaves, labeled by $x+1$ (say to the left of the root) and $n-x+1$ (to the right of the root). The set of ranked unlabeled tree shapes compatible with $\beta_x$ corresponds to the set of ranked unlabeled tree shapes with $n+1$ internal nodes ($n+2$ leaves), $x$ internal nodes for the left root subtree, and $n-x$ internal nodes for the right root subtree. 

We consider an increasing sequence of values of $n$. Supposing $n > 2x$ so that the root subtrees of $\beta_x$ cannot have the same sample size, we apply Proposition \ref{prop10}, finding that the number of ranked unlabeled tree shapes compatible with $\beta_x$ is 
\begin{equation}\label{pippo}
 {{n}\choose{x}} e_x e_{n-x}, 
\end{equation}
where $e_i$ is the number of ranked unlabeled tree shapes with $i$ internal nodes.  Following eq.~\ref{eq:res1}, the integer $e_i$ is the $i$th Euler number, $e_i=R_{i+1}$. 

The exponential generating function of the sequence $(e_i)$ is \citep{brent2013fast}
\begin{equation}\label{pino} 
\sum_{i=0}^{\infty} \frac{e_i z^i}{i!} = \sec(z) + \tan(z). 
\end{equation}
We can write the ratio $q_i=\frac{e_i}{i!}$ as 
\nocite{flajolet2009analytic}
\nocite{brent2013fast}
(Flajolet and Sedgewick 2009, p.~269; Brent and Harvey, 2013) 
\begin{equation}\label{kio}
q_i = \left\{
  \begin{array}{l l}
     2 \left( \frac{2}{\pi} \right)^{i+1}  \sum_{k=0}^{\infty}\frac{(-1)^{k}}{(2k+1)^{i+1}} , & \text{if } i \text{ is even}  \\
     2 \left[ \left( \frac{2}{\pi} \right)^{i+1} - \left( \frac{1}{\pi} \right)^{i+1} \right] \sum_{k=1}^{\infty} \frac{1}{k^{i+1}} , & \text{if } i \text{ is odd}.  \\
  \end{array} \right.
\end{equation}
As $i$ becomes large, by applying singularity analysis to  eq.~\ref{pino}, or by computing directly from eq.~\ref{kio}, we have the asymptotic relation 
\begin{equation}\label{cocco}
q_i \sim 2 \left(  \frac{2}{\pi} \right)^{i+1}.
\end{equation}

With $q_x = e_x/x!$, we rewrite eq.~\ref{pippo} as 
$n! \, q_x q_{n-x}$. Letting $n \rightarrow \infty$ for a fixed $x$, we can use eq.~\ref{kio} to rewrite $q_x$, and because $x$ is constant as $n$ grows, we can use eq.~\ref{cocco} for the asymptotic value of $q_{n-x}$. Hence, for increasing values of $n$, the number of ranked tree shapes compatible with the perfect phylogeny $\beta_x$ behaves asymptotically like the product of $n!$ and
%Rewriting (\ref{pippo}) as 
%$$\frac{e_x}{x!} \cdot \frac{e_{n-x}}{(n-x)!} \cdot n! = q_x \cdot q_{n-x} \cdot n!,$$ 
%the value of $x$ associated with the perfect phylogeny $\beta_x$ with the largest number of compatible unlabeled histories corresponds to the value that maximizes the product 
%$q_x q_{n-x}$. 
%For $n$ large enough we can use (\ref{cocco}) to write $q_{n-x}$, and by using (\ref{kio}) for $q_x$ the considered  product satisfies
\begin{equation}\label{asino}
q_x q_{n-x} \sim 4 \left( \frac{2}{\pi} \right)^{n+2} c_x,
\end{equation}
where
\begin{equation}\label{casa}
c_x = \left\{
  \begin{array}{l l}
     \sum_{k=0}^{\infty}\frac{(-1)^{k}}{(2k+1)^{x+1}} , & \text{if } x \text{ is even}  \\
      \left( 1 - \frac{1}{2^{x+1}} \right) \sum_{k=1}^{\infty} \frac{1}{k^{x+1}} , & \text{if } x \text{ is odd}.  \\
  \end{array} \right.
\end{equation}

Note that $\zeta(s) = \sum_{k=1}^{\infty} \frac{1}{k^s}$  is the Riemann zeta function. If $x$ is even, then 
$$c_x = 1 + \left(- \frac{1}{3^{x+1}} + \frac{1}{5^{x+1}} \right) + \left(- \frac{1}{7^{x+1}} + \frac{1}{9^{x+1}} \right) + ... \leq 1.$$ Among odd values of $x$, we have $c_1= \frac{3}{4} \, \zeta(2) = \pi^2/8 \approx 1.2337$ for $x=1$. For odd $x\geq 3$, we have
$$c_x < \zeta(x+1) \leq \zeta(3) \approx 1.2021 < c_1.$$
Hence, $c_1 > 1$ exceeds $c_x$ both for even $x$ and for all odd $x \geq 3$.
%increasing the value of $x$ odd, $c_x$ decreases since it can be written as 
%\begin{eqnarray}\nonumber
%c_x &=& \left( 1 - \frac{1}{2^{x+1}} \right) \cdot \sum_{k=1}^{\infty} \frac{1}{k^{x+1}} = \sum_{k=1}^{\infty} \frac{1}{k^{x+1}} - \sum_{k=1}^{\infty} \frac{1}{(2k)^{x+1}}=
%\sum_{k=1}^{\infty} \frac{1}{k^{x+1}} - \sum_{k=2,4,6,...} \frac{1}{k^{x+1}} \\\nonumber
%&=& \sum_{k=1}^{\infty} \frac{1}{k^{x+1}} - \frac{1}{2}\left(   \sum_{k=1}^{\infty} \frac{1}{k^{x+1}} - \sum_{k=1}^{\infty} \frac{(-1)^k}{k^{x+1}} \right) = \frac{1}{2}\left(   \sum_{k=1}^{\infty} \frac{1}{k^{x+1}} + \sum_{k=1}^{\infty} \frac{(-1)^k}{k^{x+1}} \right), 
%\end{eqnarray}
%where both the sums $\sum_{k=1}^{\infty} \frac{1}{k^{x+1}}$ and $\sum_{k=1}^{\infty} \frac{(-1)^k}{k^{x+1}}$ are decreasing for increasing $x$. In particular, in the latter sum 
%$$\sum_{k=1}^{\infty} \frac{(-1)^k}{k^{x+1}} = - 1 + \left(    \frac{1}{2^{x+1}} - \frac{1}{3^{x+1}} \right) + \left(   \frac{1}{4^{x+1}} - \frac{1}{5^{x+1}} \right) + ...$$ 
%each positive term 
%$$t_i(x)=\frac{1}{i^{x+1}} - \frac{1}{(i+1)^{x+1}}$$
%is a decreasing function of $x \geq 1$.

Because $c_x$ has its maximum at $x=1$, from eq.~\ref{asino}, we conclude that the product $q_x q_{n-x}$ grows asymptotically fastest for $x=1$. In particular, as $n \rightarrow \infty$, the value of $x$ for which the binary perfect phylogeny $\beta_x$ has the largest number of compatible ranked unlabeled tree shapes is $x=1$---that is, when $\beta_x = \beta_1=(2,n)$. 
%Finally, we observe that not only $\beta_1$ possesses the largest number of unlabeled histories among the binary perfect phylogenies of the form $(x+1,n-x+1)$, but it also has the largest number of histories among all the binary perfect phylogenies with $n+1$ internal nodes. Indeed, each binary perfect phylogeny $\beta$, with $x$ internal nodes for the left root subtree and $n-x$ internal nodes for the right root subtree, has a number of compatible unlabeled histories that is smaller than or equal to the number of histories compatible with $\beta_x$.   
\end{proof}

In Table 1, we can observe an example of Proposition \ref{prop:asymptotic}. The value of $b_2(10)$, or 2176, exceeds the values of $b_x(10)$ for all other values of $x$ (with the trivial exception that $b_2(10)=b_8(10)$). The asymptotic approximation from eq.~\ref{asino} gives 
\begin{equation*}
    b_2(n) \sim 2 \bigg(\frac{2}{\pi}\bigg)^{n-2} (n-2)!,
\end{equation*}
which, for $n=10$, yields $b_2(10) \approx 2175.66$.

\subsection{Ranked labeled tree shapes compatible with a labeled binary perfect phylogeny} 
\label{sec:labeled}

Propositions \ref{prop6}, \ref{prop7}, \ref{prop9} and \ref{prop10} provide recursive formulas for enumerating unranked unlabeled tree shapes and  ranked unlabeled tree shapes compatible with a binary perfect phylogeny. In these cases, a perfect phylogeny representation does not use individual sequence labels; the labels of the tips of the perfect phylogeny are simply counts of numbers of sequences. We now consider \textbf{labeled perfect phylogenies} that partition the set of labeled individual sequences. We still use the parenthetical notation described in Section \ref{sec:data} to denote a labeled perfect phylogeny, for example $\pi=(2,3)$, however, it must be understood that this labeled perfect phylogeny partitions the sampled sequences into two different sets of labeled sequences. 

\begin{figure}
  \begin{center}
\includegraphics[scale=0.5]{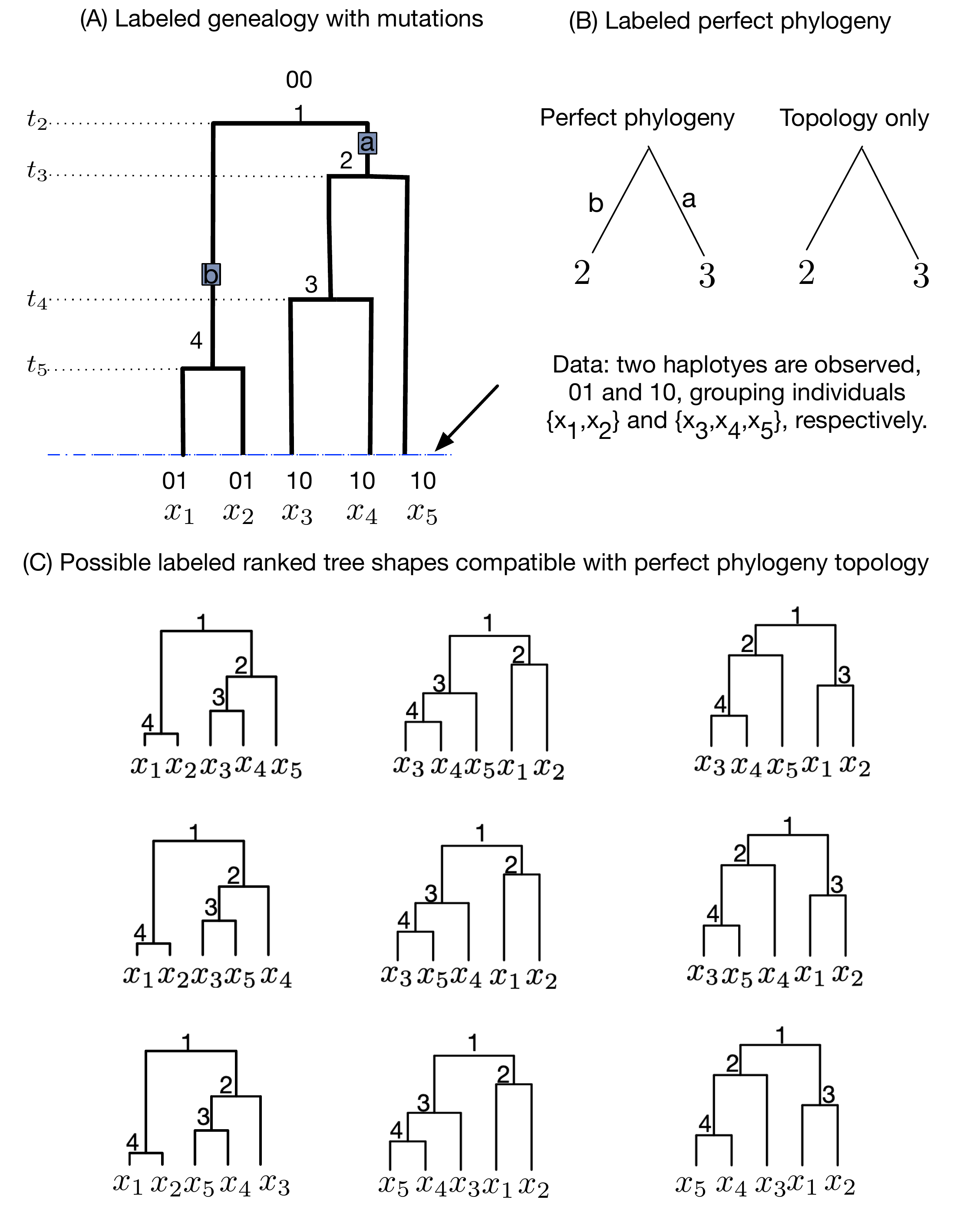}
%  % \includegraphics[width=1.0\textwidth]{Figures/tree_s.pdf}
  \end{center}
\caption{\small{\textbf{Coalescent and infinitely-many-sites generative model of binary molecular data.} \textbf{(A)} A genealogy of 5 individuals, with 2 superimposed mutations depicted as gray squares. The root is labeled by the ancestral type $00$, and the leaves are labeled by the genetic type at each of three mutated sites. The first two leaves from left to right are labeled $01$ because one mutation occurs in their path to the root.  The third, fourth and fifth individuals have one mutation in their path to the root and are labeled $10$. The order and label of the mutations is unimportant; however, individual labels $x_{1},x_{2},x_{3},x_{4},x_{5}$ are important. For ease of exposition, we label the mutations a, b. The first site corresponds to mutation a, and the second to b. \textbf{(B)} Left, a labeled perfect phylogeny representation of the observed data at the tips of (A). Data consist of 2 unique haplotypes $01$ and $10$, with frequencies 2 and 3, respectively. The corresponding frequencies are the labels of tips of the perfect phylogeny; however, it is understood that the two leaves correspond to $\{x_{1},x_{2}\}$ and $\{x_{3},x_{4},x_{5}\}$ respectively. Right, perfect phylogeny topology obtained by removing the edge labels of the perfect phylogeny. \textbf{(C)} The nine ranked labeled tree shapes compatible with the labeled perfect phylogeny topology in (B). Note that in (C), if we ignore the branching order and drop the internal node labels, in each row, the three trees are equivalent---so that each row corresponds to one of the three {\it unranked} labeled tree shapes compatible with the labeled perfect phylogeny topology in (B).}}
\label{fig:labeled_data0}
\end{figure}

Consider $\{x_{1},x_{2}\}$ and $\{x_{3},x_{4},x_{5}\}$ in the perfect phylogeny of Figure \ref{fig:labeled_data0}B. We are now interested in calculating the number of ranked labeled tree shapes compatible with a labeled binary perfect phylogeny. Figure \ref{fig:labeled_data0}C shows all the ranked labeled tree shapes compatible with the labeled perfect phylogeny. For ranked labeled tree shapes, the enumeration follows a simple recursive expression.

\begin{defn}
\label{def:labeled_comp}
\textbf{Ranked labeled tree shape $T^{L}_{n}$ compatible with a labeled perfect phylogeny $\pi \in \Pi^{L}_{n}$}. A ranked labeled tree shape with $n$ leaves, $T^{L}_{n}$, is compatible with a perfect phylogeny $\pi \in \Pi^{L}_n$ if the unranked unlabeled tree shape $T_{n}$ obtained by removing the ranks and the labels from ${T}^{L}_{n}$ is compatible with $\pi$ and the one-to-one correspondence between the $k$ leaves of $\pi$ and the $k$ disjoint subtrees of $T^{L}_{n}$ correspond to the same partition of the individual sequences.
\end{defn}

%\begin{defn}
%\label{def:labeled_comp}
%\textbf{Ranked labeled tree shape $T^{L}_{n}$ compatible with %a perfect phylogeny $\pi \in \Pi_{n}$}. A ranked labeled tree shape with $n$ leaves, $T^{L}_{n}$, is compatible with a perfect phylogeny $\pi \in \Pi_{n}$  if the unranked unlabeled tree shape $T_{n}$ obtained by removing the rankings and the labels from ${T}^{L}_{n}$ is compatible with $\pi$.
%\end{defn}

\begin{prop} For $n_1, n_2 \geq 1$ and $\pi_{1}\in \Pi_{n_{1}}^L, \pi_{2}\in \Pi_{n_{2}}^L$ the number of ranked labeled tree shapes compatible with a labeled binary perfect phylogeny $\pi=(\pi_{1},\pi_{2})$ is
%with $n$ leaves $|\mathcal{G}^{L}_{c}(\pi)|$, such that $\pi_{1}\in \Pi_{n_{1}}$ and $\pi_{2}\in \Pi_{n_{2}}$, $n=n_{1}+n_{2}$ is
\begin{equation}
|\mathcal{G}^{L}_{c}(\pi)|=
\binom{n_{1}+n_{2}-2}{n_{1}-1}|\mathcal{G}^{L}_{c}(\pi_{1})| \, |\mathcal{G}^{L}_{c}(\pi_{2})|.
\end{equation}
\label{prop13}
\end{prop}
\begin{proof}
%By Definitions \ref{def:treeshape_comp}, \ref{def:ranked_comp}, and \ref{def:labeled_comp}, 
We can count the number of ranked labeled tree shapes by dividing $\pi$ at the root into two subtrees, one with $n_1$ leaves and perfect phylogeny $\pi_1$, and the other with $n_2$ leaves and perfect phylogeny $\pi_2$, both partitioning the sampled sequences. The number of such trees is the product of the numbers of ranked labeled trees for the two subtrees and the number of ways of interleaving the internal nodes of the two subtrees. In this case, the two perfect phylogenies $\pi_{1}$ and $\pi_{2}$ can never be identical because they correspond to different sets of sequences.
\end{proof}

%%%%%%%%%%%%%%%%%%%%%%%% Figure 8 %%%%%%%%%%%%%%%%%%%%%%%%%%%
%%%%%%%%%%%%%%%%%%%%%%%%%%%%%%%%%%%%%%%%%%%%%%%%%%%%%%%%%%%%%

Counts for the number of ranked labeled tree shapes for some of the perfect phylogenies of 10 samples (with an arbitrary labeling) appear in Table 1. Given a perfect phylogeny in the table, we can observe that the number of ranked labeled tree shapes far exceeds the number of ranked unlabeled tree shapes.

Continuing with ((4,2),6), the number of ranked labeled tree shapes compatible with this (arbitrarily labeled) perfect phylogeny is ${10 \choose 5} |\mathcal{G}_c^L((4,2))| \, |\mathcal{G}_c^L((6))| = {10 \choose 5} {4 \choose 3} \, |\mathcal{G}_c^L((4))|\, |\mathcal{G}_c^L((2))| \,|\mathcal{G}_c^L((6))| =  {10 \choose 5} {4 \choose 3} L_4 L_2 L_6 = 252 \times 4 \times 18 \times 1 \times 2700=48,988,800$.

We can obtain a result analogous to Proposition \ref{prop:asymptotic}; we characterize, for binary labeled perfect phylogenies $B_x(n)=(x,n-x)$, the one compatible with the largest number of ranked labeled tree shapes. Let $b_x'(n)$ denote the number of ranked labeled tree shapes compatible with $B_x(n)$. 
\begin{prop}
\label{prop:max}
Fix $n \geq 2$. Among the values $b_1'(n), b_2'(n), \ldots, b_{\lfloor \frac{n}{2} \rfloor}'(n)$, the largest is $b_1'(n)$.
\end{prop}
\begin{proof}
Applying Proposition \ref{prop13} , we have $b_x'(n)={n-2 \choose x-1} \, L_x \, L_{n-x}$. Simplifying with eq.~\ref{eq:Ln}, we obtain
$b_x'(n) = [n! \, (n-2)! / {2^{n-2}}]{n \choose x}^{-1}$.
As it is quickly verified that the binomial coefficients ${n \choose x}$ increase monotonically from $x=1$ to $x=\lfloor \frac{n}{2} \rfloor$, $b_x'$ 
decreases monotonically from $x=1$ to  $x=\lfloor \frac{n}{2} \rfloor$.
\end{proof}
An example of Proposition \ref{prop:max} is visible in Table 1, in which $b_1'(10)=57,153,600$ exceeds $b_2'(10)$, $b_3'(10)$, $b_4'(10)$, and $b_5'(10)$.

%In the next section, we consider the enumeration for the case of multifurcating perfect phylogeny topologies. This calculation relies on finding all the distinct binary perfect phylogeny topologies and summing the cardinalities obtained from each binary perfect phylogeny. 

%%%%%%%%%%%%%%%%%%%%%%%%%%%%%%%%%%%%%%%%%%%%%%%%%%%%%%%%%%%%%

\subsection{Unranked labeled tree shapes compatible with a labeled binary perfect phylogeny}
\label{sec:unrankedlabeled}

Continuing with the labeled perfect phylogenies from Section \ref{sec:labeled}, we now count the unranked labeled binary perfect phylogenies compatible with a labeled binary perfect phylogeny.

Consider $\{x_{1},x_{2}\}$ and $\{x_{3},x_{4},x_{5}\}$ in the perfect phylogeny of Figure \ref{fig:labeled_data0}B. We calculate the number of unranked labeled tree shapes compatible with a labeled binary perfect phylogeny. Each row of Figure \ref{fig:labeled_data0}C corresponds to one of the unranked labeled tree shapes compatible with the labeled perfect phylogeny. 

\begin{defn}
\label{def:unrankedlabeled_comp}
\textbf{Ranked labeled tree shape $T^{X}_{n}$ compatible with a labeled perfect phylogeny $\pi \in \Pi^{L}_{n}$}. An unranked labeled tree shape with $n$ leaves, $T^{X}_{n}$, is compatible with a perfect phylogeny $\pi \in \Pi^{L}_n$ if the unranked unlabeled tree shape $T_{n}$ obtained by removing the labels from ${T}^{X}_{n}$ is compatible with $\pi$ and the one-to-one correspondence between the $k$ leaves of $\pi$ and the $k$ disjoint subtrees of $T^{X}_{n}$ correspond to the same partition of the individual sequences.
\end{defn}

\begin{prop} For $n_1, n_2 \geq 1$ and $\pi_{1}\in \Pi_{n_{1}}^L, \pi_{2}\in \Pi_{n_{2}}^L$, the number of ranked labeled tree shapes compatible with a labeled binary perfect phylogeny $\pi=(\pi_{1},\pi_{2})$ is
\begin{equation}
|\mathcal{G}^{X}_{c}(\pi)|=
|\mathcal{G}^{X}_{c}(\pi_{1})| \, |\mathcal{G}^{X}_{c}(\pi_{2})|.
\end{equation}
\label{prop16}
\end{prop}
\begin{proof}
We divide $\pi$ at the root into two subtrees, one with $n_1$ leaves and perfect phylogeny $\pi_1$, and the other with $n_2$ leaves and perfect phylogeny $\pi_2$. The subtrees must partition the sampled sequences in the same way as $\pi$. The number of such trees is the simply product of the numbers of unranked labeled trees for the two subtrees. As in Proposition \ref{prop13}, perfect phylogenies $\pi_{1}$ and $\pi_{2}$ are not identical because they correspond to different sets of sequences; with the ranking dropped, unlike in Proposition \ref{prop13}, we need not consider the number of ways of interleaving the internal nodes of the two subtrees.
\end{proof}

For some of the perfect phylogenies of 10 samples (with an arbitrary labeling), counts for the number of unranked labeled tree shapes appear in Table 1. The number of unranked labeled tree shapes far exceeds the number of unranked unlabeled tree shapes, and it generally exceeds the number of ranked unlabeled tree shapes.

For the example ((4,2),6), the number of unranked labeled tree shapes compatible with this (arbitrarily labeled) perfect phylogeny is $|\mathcal{G}_c^X((4,2))| \, |\mathcal{G}_c^X((6))| =  |\mathcal{G}_c^X((4))| \,|\mathcal{G}_c^X((2))| \,|\mathcal{G}_c^X((6))| =  X_4 X_2 X_6 = 15 \times 1 \times 945 =14,175$.

For binary labeled perfect phylogenies $B_x(n)=(x,n-x)$, the one compatible with the largest number of unranked labeled tree shapes follows the result of Proposition \ref{prop:max}. Let $b_x''(n)$ denote the number of unranked labeled tree shapes compatible with $B_x(n)$. 
\begin{prop}
\label{prop:max2}
Fix $n \geq 2$. Among the values $b_1''(n), b_2''(n), \ldots, b_{\lfloor \frac{n}{2} \rfloor}''(n)$, the largest is $b_1''(n)$.
\end{prop}
\begin{proof}
Applying Proposition \ref{prop16} , we have $b_x''(n)= X_x \, X_{n-x}$ for $1 \leq x \leq \lfloor \frac{n}{2} \rfloor$. Simplifying with eq.~\ref{eq:Xn}, we obtain
$$b_x''(n) = \frac{(n-2)!}{2^{n-2}} \frac{{2x-2 \choose x-1}{2n-2x-2 \choose n-x-1}}{{n-2 \choose x-1}}.$$
Then $b_{x+1}''(n)/b_{x}''(n) = \frac{2x-1}{2n-2x-3} \leq 1$ for $1 \leq x \leq \frac{n-1}{2}$, with equality requiring $x=\frac{n-1}{2}$, so that $b_{x}''(n)$ monotonically decreases from $x=1$ to $x=\lfloor \frac{n}{2} \rfloor$.
\end{proof}
In Table 1, we observe that as in Proposition \ref{prop:max2},  $b_1''(10)=2,027,025$ exceeds $b_2''(10)$, $b_3''(10)$, $b_4''(10)$, and $b_5''(10)$.

\section{Enumeration for multifurcating perfect phylogenies} \label{sec:four}
\label{sec:multifurcating}

Recall that perfect phylogenies need not be strictly binary, and that nodes can have more than two descendants. To complete the description of the numbers of trees of various types that are compatible with a perfect phylogeny, we must consider multifurcating perfect phylogenies. We proceed by reducing the multifurcating case to the binary case that has already been solved.

\begin{figure}
\centering
\includegraphics[scale=0.5]{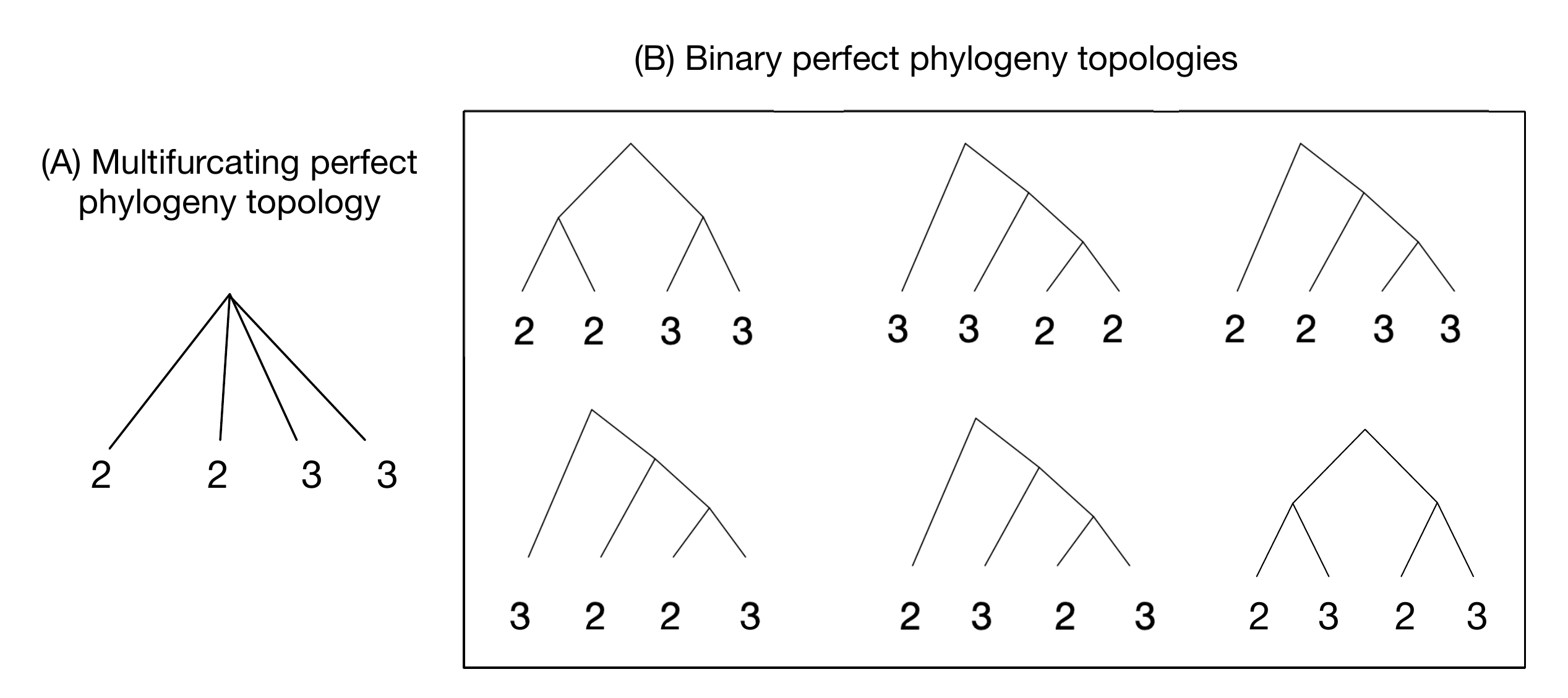}
\caption{\small{\textbf{Example of all possible binary perfect phylogeny topologies for a given multifurcating perfect phylogeny topology.} The binary perfect phylogenies are obtained from a multifurcating perfect phylogeny by resolving multifurcating nodes into sequences of bifurcations.}}
\label{fig:compatBin}
\end{figure}

We now consider a \textbf{multifurcating perfect phylogeny} that consists of a single internal node subtending $k$ leaves with labels $n_{1},n_{2},\ldots,n_{k}$. An example is depicted in Figure \ref{fig:compatBin}. Because multiple leaves can each correspond to groups with the same number of samples, so that the same numerical label can be assigned to many of those leaves, it is convenient to denote the vector of unique labels by $\mathbf{a}=(a_{1},a_{2},\ldots,a_{s})$ and the corresponding vector of their multiplicities by $\mathbf{m}=(m_{1},m_{2},\ldots,m_{s})$, where $m_{j}$ denotes the number of leaves with label $a_{j}$, $1 \leq j \leq s 
\leq k$. In the example of Figure \ref{fig:compatBin}, $\mathbf{a}=(2,3)$ and $\mathbf{m}=(2,2)$, as two leaves $(m_1=2)$ have label 2 $(a_1=2)$ and two leaves $(m_2=3)$ have label 3 $(a_2=3)$.

We extend the notion of the binary perfect phylogeny poset to the multifurcating case. We define $\pi \leq \sigma$ for two multifurcating perfect phylogenies if $\sigma$ can be obtained by sequentially collapsing pairs of pendant edges of $\pi$. Given two multifurcating perfect phylogenies $\pi_{1}$ and $\pi_{2}$, their meet $\pi_{1} \wedge \pi_{2}$ is the largest multifurcating perfect phylogeny that refines both $\pi_{1}$ and $\pi_{2}$. For example, the meet between $\pi_{1}=(1,2,3,(2,2))$ and $\pi_{2}=(1,2,2,(2,3))$ is given by:
\begin{align*}
(1,2,3,(2,2)) \wedge (1,2,2,(2,3)) &= (1,(2,2),(2,3)).
\end{align*}
Similarly, their join is the smallest multifurcating perfect phylogeny $\pi_{1} \vee \pi_{2}$ for which both $\pi_{1}$ and $\pi_{2}$ are refinements:
\begin{align*}
(1,2,3,(2,2)) \vee (1,2,2,(2,3)) &= (1,2,2,2,3).
\end{align*}

The lattice structure enables us to count the number of ranked unlabeled tree shapes compatible with a multifurcating perfect phylogeny $\pi=(n_{1},n_{2},\ldots,n_{k})$. We use a recursive inclusion-exclusion principle  %\textcolor{green}{$\mathcal{B}(\mathbf{a},\mathbf{m})$}, 
with label vector $\mathbf{a}$ and multiplicities $\mathbf{m}$. The key idea is to decompose the computation into a sum over all possible binary perfect phylogenies, applying Propositions \ref{prop9} and \ref{prop10} to each binary perfect phylogeny. To recursively generate all possible binary perfect phylogenies from $\pi$, we define the operator $\mathcal{B}_{i,j}(\pi)$ that collapses two leaves with labels $a_{i}$ and $a_{j}$ in $\pi$. For example $\mathcal{B}_{2,3}(2,2,3,4)=((2,3),2,4)$.
%(n_{1},\ldots,n_{k})=(n_{1},\ldots,n_{k})$ with multiplicities: $\mathbf{m}=(m_{1},\ldots,m_{s})$. 
%Let $\mathcal{B}(n_{1},\ldots,n_{k})=\mathcal{B}(\mathbf{a},\mathbf{m})$, then 
If $\sum^{s}_{i=1} m_{i}>2$, then
\begin{align}\label{eq:first}
|\mathcal{G}_{c}(\pi)|&=\underbrace{\sum_{i=1}^s|\mathcal{G}_{c}(\mathcal{B}_{i,i}(\pi))|\,1_{m_{i}>1}}_{\small{\substack{\text{collapsing two pendant edges}\\ \text{with the same leaf values}}}}+\underbrace{\sum_{i=1}^{s-1} \sum_{j=i+1}^s|\mathcal{G}_{c}(\mathcal{B}_{i,j}(\pi))\,|1_{m_{i}>0} \, 1_{m_{j}>0}}_{\small{\substack{\text{collapsing two pendant edges}\\ \text{with different leaf values}}}} \nonumber \\
& \quad - \underbrace{\sum_{i=1}^{s-1} \sum_{j=i+1}^s|\mathcal{G}_{c}(\mathcal{B}_{i,i}(\pi) \wedge \mathcal{B}_{j,j}(\pi))|\,1_{m_{i}>1}\,1_{m_{j}>1}}_{\small{\substack{\text{collapsing all pairs containing two distinct pairs of pendant edges,}\\ \text{each pair with the same leaf values}}}} \nonumber\\
& \quad - \underbrace{\sum_{i=1}^{s-1} \sum_{j=i+1}^s \sum_{k=1 \atop k \neq i, k \neq j}^s |\mathcal{G}_{c}(\mathcal{B}_{i,j}(\pi) \wedge \mathcal{B}_{k,k}(\pi)) | \, 1_{m_{i}>0} \, 1_{m_{j}>0} \, 1_{m_{k}>1}}_{\small{\substack{\text{collapsing a pair of edges with different leaf values}\\ \text{and collapsing a pair of edges with the same leaf values}}}} \nonumber\\
& \quad - \underbrace{\sum_{i=1}^{s-1} \sum_{j=i+1}^s \sum_{k=1 \atop k \neq i, k \neq j}^{s-1} \sum_{\ell=k+1 \atop \ell \neq i, \ell \neq j}^s  |\mathcal{G}_{c}(\mathcal{B}_{i,j}(\pi) \wedge \mathcal{B}_{k,\ell}(\pi))|\, 1_{m_{i}>0}\, 1_{m_{j}>0}\, 1_{m_{k}>0}\, 1_{m_{\ell}>0}}_{\small{\substack{\text{collapsing two different pairs of pendant edges,}\\ \text{each pair with different leaf values}}}}. 
\end{align}
To interpret eq.~\ref{eq:first} as an inclusion-exclusion formula, notice that the first two sums that are added on the right-hand side of eq.~\ref{eq:first} correspond to enumerations of single events (so that the sum is analogous to a union $\cup A_{i}$), and the following three sums that are subtracted correspond to intersections of pairs of these events (analogous to intersections $A_{i} \cap A_{j}$). 

Eq.~\ref{eq:first} provides a recursive approach for counting the number of ranked unlabeled tree shapes compatible with a multifurcating perfect phylogeny by expressing the calculation in terms of binary perfect phylogenies. The recursive application of the equation proceeds until all terms reach $\sum^{s}_{i=1}m_{i}=2$, when the binary perfect phylogenies are reached. 
% Formerly example 3
\begin{examp}
The number of ranked unlabeled tree shapes compatible with $\pi=(2,2,3,3)$ is:
\begin{align*}
|\mathcal{G}^{T}_{c}(2,2,3,3)| & =|\mathcal{G}^{T}_{c}((2,2),3,3)| +|\mathcal{G}^{T}_{c}(2,2,(3,3))| +|\mathcal{G}^{T}_{c}((2,3),2,3)| -|\mathcal{G}^{T}_{c}((2,2),(3,3))|\\
& = \big[ |\mathcal{G}^{T}_{c}((2,2),(3,3))| +|\mathcal{G}^{T}_{c}(((2,2),3),3)| \big] + \big[ |\mathcal{G}^{T}_{c}((2,2),(3,3))| +|\mathcal{G}^{T}_{c}(((3,3),2),2)| \big]\\
& \quad + \big[ |\mathcal{G}^{T}_{c}(((2,3),2),3)| +|\mathcal{G}^{T}_{c}(((2,3),3),2)| + |\mathcal{G}^{T}_{c}((2,3),(2,3))| \big] -|\mathcal{G}^{T}_{c}((2,2),(3,3))| \\
& =|\mathcal{G}^{T}_{c}((2,2),(3,3))| +|\mathcal{G}^{T}_{c}(((2,2),3),3)| +|\mathcal{G}^{T}_{c}(((3,3),2),2)| +|\mathcal{G}^{T}_{c}(((2,3),2),3)| \\
& \quad  +|\mathcal{G}^{T}_{c}(((2,3),3),2)| + |\mathcal{G}^{T}_{c}((2,3),(2,3))| \\
& =168+280+144+420+360+315=1687.
\end{align*}
In obtaining this sum, in intermediate steps, we use the fact that the values of $\mathcal{G}_{c}^T$ for (2), (3), (2,2), (3,3), (2,3), ((2,2),3), ((3,3),2), ((2,3),2)), ((2,3),3), and  are 1, 1, 1, 3, 3, 10, 18, 15, and 45, respectively.
\end{examp}

For counting the number of unranked unlabeled tree shapes compatible with $\pi=(n_{1},n_{2},\ldots,n_{k})$, we simply replace $\mathcal{G}^{T}_{c}$ with $\mathcal{G}_{c}$ in eq.~\ref{eq:first}. We use Propositions \ref{prop6} and \ref{prop7} in place of Propositions \ref{prop9} and \ref{prop10}.
\begin{examp} The number of unranked unlabeled tree shapes compatible with $\pi=(2,2,3,3)$ is:
\begin{align*}
|\mathcal{G}_{c}(2,2,3,3)| &= |\mathcal{G}_{c}((2,2),(3,3))|+|\mathcal{G}_{c}(((2,2),3),3)|+|\mathcal{G}_{c}(((3,3),2),2)|\\
 & \quad +|\mathcal{G}_{c}(((2,3),2),3)|+|\mathcal{G}_{c}(((2,3),3),2)|+ |\mathcal{G}_{c}((2,3),(2,3))| \\
 &=1+1+1+1+1+1=6.
\end{align*}
This example is quite straightforward; the values of $\mathcal{G}_{c}$ for the perfect phylogenies that appear in intermediate steps---(2), (3), (2,2),  (3,3), (2,3), ((2,2),3), ((3,3),2), ((2,3),2)), and ((2,3),3)---all equal 1.
\end{examp}

To count the number of ranked labeled tree shapes compatible with a labeled multifurcating perfect phylogeny $\pi=(n_{1},n_{2},\ldots,n_{k})$, we assume that although any leaf in the perfect phylogeny can have multiplicity larger than one, each leaf is uniquely defined by its associated samples, all of which are all assumed to have different labels. Therefore, we take $\mathbf{a}=(n_{1},n_{2},\ldots,n_{k})$ and $\mathbf{m}=(1,1,\ldots,1)$. Eq.~\ref{eq:first} reduces to
\begin{align}\label{eq:first_2}
|\mathcal{G}^{L}_{c}(\pi)|&=\underbrace{\sum_{i=1}^{s-1} \sum_{j=i+1}^s|\mathcal{G}^{L}_{c}(\mathcal{B}_{i,j}(\pi))|\, 1_{m_{i}>0}\, 1_{m_{j}>0}}_{\small{\substack{\text{collapsing two pendant edges}}}} \nonumber \\
& \quad - \underbrace{\sum_{i=1}^{s-1} \sum_{j=i+1}^s \sum_{k=1 \atop k \neq i, k \neq j}^{s-1} \sum_{\ell=k+1 \atop \ell\neq i, \ell \neq j}^s|\mathcal{G}^{L}_{c}(\mathcal{B}_{i,j}(\pi) \wedge \mathcal{B}_{k,\ell}(\pi))|\, 1_{m_{i}>0}\, 1_{m_{j}>0}\, 1_{m_{k}>0}\, 1_{m_{\ell}>0}}_{\small{\substack{\text{collapsing two pairs of pendant edges}}}}.
\end{align}
The enumeration makes use of Proposition \ref{prop13}. 

\begin{examp}
Consider a labeled multifurcating perfect phylogeny that groups $2$, $2$, $3$, and $3$ samples at the root. We assume that $\mathbf{a}=(a_{1},a_{2},a_{3},a_{4})=(2,2,3,3)$. Applying the recursion formula in eq.~\ref{eq:first_2}, we get
\begin{align*}
|\mathcal{G}^{L}_{c}(a_{1},a_{2},a_{3},a_{4})| &= |\mathcal{G}^{L}_{c}((a_{1},a_{2}),a_{3},a_{4})|+|\mathcal{G}^{L}_{c}((a_{1},a_{3}),a_{2},a_{4})|+|\mathcal{G}^{L}_{c}((a_{1},a_{4}),a_{2},a_{3})|\\
& \quad + |\mathcal{G}^{L}_{c}((a_{2},a_{3}),a_{1},a_{4})|+|\mathcal{G}^{L}_{c}((a_{2},a_{4}),a_{1},a_{3})|+|\mathcal{G}^{L}_{c}((a_{3},a_{4}),a_{1},a_{2})|\\
& \quad - |\mathcal{G}^{L}_{c}((a_{1},a_{2}),(a_{3},a_{4}))|-|\mathcal{G}^{L}_{c}((a_{1},a_{3}),(a_{2},a_{4}))|-|\mathcal{G}^{L}_{c}((a_{1},a_{4}),(a_{2},a_{3}))|\\
 &=|\mathcal{G}^{L}_{c}((2,2),3,3)|+4|\mathcal{G}^{L}_{c}((2,3),2,3)|+|\mathcal{G}^{L}_{c}((3,3),2,2)|\\
 & \quad -|\mathcal{G}^{L}_{c}((2,2),(3,3))|-2|\mathcal{G}^{L}_{c}((2,3),(2,3))|.
 %&=6048+5040+2592+3780+3240=20700.
\end{align*}
Now, because 
\begin{align*}
|\mathcal{G}^{L}_{c}(a_{1},a_{2},a_{3})| &= |\mathcal{G}^{L}_{c}((a_{1},a_{2}),a_{3})|+|\mathcal{G}^{L}_{c}((a_{1},a_{3}),a_{2})|+|\mathcal{G}^{L}_{c}((a_{2},a_{3}),a_{1})|,
\end{align*}
we have
\begin{align*}
|\mathcal{G}^{L}_{c}((2,2),3,3)| &= 2|\mathcal{G}^{L}_{c}(((2,2),3),3)|+|\mathcal{G}^{L}_{c}((2,2),(3,3))| \\
|\mathcal{G}^{L}_{c}((2,3),2,3)| &= |\mathcal{G}^{L}_{c}(((2,3),2),3)|+|\mathcal{G}^{L}_{c}(((2,3),3),2)|+|\mathcal{G}^{L}_{c}((2,3),(2,3))| \\
|\mathcal{G}^{L}_{c}((3,3),2,2)| &= 2|\mathcal{G}^{L}_{c}(((3,3),2),2)|+|\mathcal{G}^{L}_{c}((2,2),(3,3))|.
\end{align*}
Summing all terms, we get
\begin{align*}
|\mathcal{G}^{L}_{c}(a_{1},a_{2},a_{3},a_{4})| &=2|\mathcal{G}^{L}_{c}(((2,2),3),3)|+2|\mathcal{G}^{L}_{c}(((3,3),2),2)|+|\mathcal{G}^{L}_{c}((2,2),(3,3))|\\
& \quad +4|\mathcal{G}^{L}_{c}(((2,3),2),3)|+4|\mathcal{G}^{L}_{c}(((2,3),3),2)|+2|\mathcal{G}^{L}_{c}((2,3),(2,3))|\\
&=2\times 5040 + 2 \times 2592 + 6048 + 4 \times 3780 + 4 \times 3240 + 2 \times 5670= 60,732.
\end{align*}
In obtaining this sum, we use the fact that the values of $\mathcal{G}_{c}^L$ for (2), (3), (2,2), (3,3), (2,3), ((2,2),3), ((3,3),2), ((2,3),2)), and ((2,3),3), and  are 1, 3, 2, 54, 9, 60, 324, 45, and 405, respectively.
\end{examp}

The number of unranked labeled tree shapes compatible with $\pi=(n_{1},n_{2},\ldots,n_{k})$ is obtained by replacing $\mathcal{G}^{L}_{c}$ with $\mathcal{G}^{X}_{c}$ in eq.~\ref{eq:first_2}. We use Proposition \ref{prop16} in place of Proposition \ref{prop13}.
\begin{examp} The number of unranked labeled tree shapes compatible with a labeled multifurcating perfect phylogeny that groups 2, 2, 3, and 3 samples at the root, with $\mathbf{a}=(a_{1},a_{2},a_{3},a_{4})=(2,2,3,3)$ is:
\begin{align*}
|\mathcal{G}^{X}_{c}(a_{1},a_{2},a_{3},a_{4})| &=2|\mathcal{G}^{X}_{c}(((2,2),3),3)|+2|\mathcal{G}^{X}_{c}(((3,3),2),2)|+|\mathcal{G}^{X}_{c}((2,2),(3,3))|\\
& \quad +4|\mathcal{G}^{X}_{c}(((2,3),2),3)|+4|\mathcal{G}^{X}_{c}(((2,3),3),2)|+2|\mathcal{G}^{X}_{c}((2,3),(2,3))|\\
&=2\times 9 + 2 \times 9 + 9 + 4 \times 9 + 4 \times 9 + 2 \times 9= 135.
\end{align*}
The sum uses values of $\mathcal{G}_{c}^{X}$ for (2), (3), (2,2), (3,3), (2,3), ((2,2),3), ((3,3),2), ((2,3),2)), and ((2,3),3), equal to 1, 3, 1, 9, 3, 3, 9, 3, and 9, respectively.
\end{examp}

%%%%%%%%%%%%%%%%%%%%%%%%%%%%%%%%%%%%%%%%% Figure 9 %%%%%%%%%%%%%%%%%%%
%%%%%%%%%%%%%%%%%%%%%%%%%%%%%%%%%%%%%%%%%%%%%%%%%%%%%%%%%%%%%%%%%%%%%%

%Table 1 shows the number of unranked unlabeled tree shapes, ranked unlabeled tree shapes, and ranked labeled tree shapes of some perfect phylogenies in $\Pi_{10}$. 
%Figure \ref{fig:tree_count2} shows the number of ranked and labeled tree shapes for the same perfect phylogenies. 
%The reduction in the state space of trees by removing the labels (ranked and labeled versus ranked tree shapes) is remarkable.

%\begin{figure}
 % \begin{center}
%\includegraphics[scale=0.8]{Figures/results.pdf}
% \end{center}
% \vspace{-1cm}
%\caption{\small{\textbf{Number of ranked tree shapes (first row) and number of tree shapes (second row) for perfect phylogenies with two tips (first column) and three tips (second column) for $n=10$.} }}
%\label{fig:tree_count}
%\end{figure}

%\begin{figure}
%  \begin{center}
%\includegraphics[scale=0.45]{Figures/n10_labeled.pdf}
 %\end{center}
 %\vspace{-1cm}
%\caption{\small{\textbf{Number of ranked and labeled tree shapes for perfect phylogenies with two tips (left) and three tips (right) for $n=10$.} }}
%\label{fig:tree_count2}
%\end{figure}

%\textcolor{red}{The number of ranked tree shapes with $c$ cherries is }

%\begin{theorem}
%The maximum number of ranked tree shapes of $n$ leaves compatible with a perfect phylogeny occurs when the perfect phylogeny is a multifurcating phylogeny of degree $n$ or $n-1$.
%\end{theorem}

% latex table generated in R 4.0.3 by xtable 1.8-4 package
% Wed Feb 24 18:06:07 2021
\begin{table}[ht]
\label{table:tree_count}
\begin{center}
\caption{Number of trees compatible with example perfect phylogenies of 10 samples.}
\begin{tabular}{rrrrr}
\hline
Perfect & Unranked unlabeled & Ranked unlabeled & Ranked labeled & Unranked labeled \\ 
phylogeny & tree shapes & tree shapes & tree shapes & tree shapes \\  \hline
(9,1)     & 46 & 1385 & 57,153,600 & 2,027,025 \\ 
(8,2)     & 23 & 2176 & 12,700,800 & 135,135 \\ 
(7,3)     & 11 & 1708 & 4,762,800  & 31,185 \\ 
(6,4)     & 12 & 1792 & 2,721,600  & 14,175 \\ 
(5,5)     & 6  & 875  & 2,268,000  & 11,025 \\ 
((8,1),1) & 23 & 272  & 1,587,600  & 135,135 \\ 
((7,2),1) & 11 & 427  & 396,900 & 10,395  \\ 
((6,3),1) & 6  & 336  & 170,100 & 2835 \\ 
((5,4),1) & 6  & 350  & 113,400 & 1575 \\ 
((7,1),2) & 11 & 488  & 453,600 & 10,395 \\ 
((6,2),2) & 6  & 768  & 129,600 & 945 \\ 
((5,3),2) & 3  & 600  & 64,800  & 315 \\ 
((4,4),2) & 3  & 320  & 51,840  & 225 \\ 
((6,1),3) & 6  & 448  & 226,800 & 2835 \\ 
((5,2),3) & 3  & 700  & 75,600  & 315 \\ 
((4,3),3) & 2  & 560  & 45,360  & 135 \\ 
((5,1),4) & 6  & 560  & 181,440 & 1575 \\ 
((4,2),4) & 4  & 896  & 72,576  & 225 \\ 
((3,3),4) & 2  & 336  & 54,432  & 135 \\ 
((4,1),5) & 5  & 560  & 226,800 & 1575 \\ 
((3,2),5) & 3  & 735  & 113,400 & 315 \\ \hline
\end{tabular}
\end{center}
The entries in the table are obtained by repeated use of Propositions \ref{prop6} and \ref{prop7} for unranked unlabeled tree shapes, \ref{prop9} and \ref{prop10} for ranked unlabeled tree shapes, \ref{prop13} for ranked labeled tree shapes, and \ref{prop16} for unranked labeled tree shapes. An arbitrary labeling of the perfect phylogeny is assumed for counting the associated ranked and unranked labeled tree shapes.
\end{table}

\renewcommand{\algorithmicrequire}{\textbf{Input:}} 
\renewcommand{\algorithmicensure}{\textbf{Output:}}

%TODO: Change algorithm 1 to group nodes according to size (and symmetries)

%\textcolor{red}{Aim for Journal of Mathematical Biology}

\section{Conclusion}

The infinitely-many-sites mutations model is a popular model of molecular variation for problems of population genetics \citep{wakeley_coalescent_2008} and related areas \citep{jones2020inference}, in which constraints are imposed on the space of trees that can explain the observed patterns of molecular variation. A realization of the coalescent model on a genealogy and a superimposed infinitely-many-sites mutation model can be summarized as a perfect phylogeny. Here, we have examined combinatorial properties of the genealogical tree structures that are compatible with a perfect phylogeny, demonstrating that the binary perfect phylogenies possess a lattice structure (Theorem \ref{thm:lattice}). We have used this lattice structure to provide recursive enumerative results counting the trees---unranked unlabeled trees, ranked unlabeled trees, ranked labeled trees, and unranked labeled trees---compatible with binary and multifurcating perfect phylogenies.

In our enumerative results, the count of the number of trees of a specified type that are compatible with a perfect phylogeny is obtained by a decomposition of the perfect phylogeny at its root. The number of associated trees is obtained by counting trees for each subtree immediately descended from the root of the perfect phylogeny---and where appropriate, counting interleavings of nodes within those trees, taking care to consider cases that avoid double-counting, or both. This same technique was applicable for each of the types of trees we considered, appearing in Sections \ref{sec:unranked}, \ref{sec:ranked}, \ref{sec:labeled}, \ref{sec:unrankedlabeled}, and \ref{sec:multifurcating}. Owing to the recursive structure of the computation, the decomposition itself proceeds rapidly from the root through the internal nodes, so that a count can be quickly obtained even if the number itself is large.

We obtained results concerning the cherry perfect phylogenies with the largest numbers of ranked unlabeled, unranked labeled, and ranked labeled tree shapes (Propositions \ref{prop:asymptotic}, \ref{prop:max}, and \ref{prop:max2}), and it will be informative to seek a similar result for the unranked unlabeled case. The result in Proposition \ref{prop:asymptotic} on asymptotic growth of the number of ranked unlabeled tree shapes compatible with a binary perfect phylogeny is reminiscent of a result concerning ``lodgepole'' trees. A number of studies have examined another combinatorial structure for evolutionary trees, the number of ``coalescent histories'' associated with a labeled species tree and its matching labeled gene tree. These coalescent histories encode different evolutionary scenarios possible for the coalescence of gene lineages on a species tree. \citet{disanto2015coalescent} found that the lodgepole trees, a class of trees in which cherry nodes with 2 descendants successively branch from a single species tree edge, possesses a particularly large number of coalescent histories. Similarly, in Proposition \ref{prop:asymptotic}, as $n$ increases, the number of ranked unlabeled tree shapes compatible with a cherry perfect phylogeny is largest when the perfect phylogeny has one subtree with sample size 2.
%\nocite{DisantoAndRosenberg15}

Perfect phylogenies have been widely studied in varied estimation problems, for the ``perfect phylogeny problem'' asking whether a perfect phylogeny can be constructed from data given on a set of characters \citep{agarwala1994faster, kannan1997fast, felsenstein2004inferring,gusfield2014recombinatorics, steel16}, statistical inference of evolutionary parameters under the coalescent \citep{griffiths_sampling_1994,StephensDonnelly2000,TavareNotes,Palacios2019,cappello2020tajima}, and algorithmic estimation of haplotype phase from diploid data \citep{gusfield2002haplotyping,bafna2004note,gusfield2014recombinatorics}. However, the literature on perfect phylogenies has largely focused on such applications and on algorithmic problems of obtaining perfect phylogenies from data under various constraints, with little emphasis on the enumerative combinatorics of the perfect phylogenies themselves, and of their associated refinements. In describing a lattice for the binary perfect phylogenies with sample size $n$, this study suggests that the mathematical properties of sets of perfect phylogenies as combinatorial structures {\it per se} can be informative. The link to coalescent histories suggests possible connections to related concepts such as ``ancestral configurations'' \citep{wu2012coalescent,disanto2017enumeration}, which also can be described in terms of lattices (E.~Alimpiev \& N.A.R., unpublished); it will be useful to consider perfect phylogenies alongside such structures arising in the combinatorics of evolutionary trees.

Finally, returning to considerations of coalescent-based inference from sequences, recall that inference of evolutionary parameters from a given perfect phylogeny is performed by integrating over the space of genealogies. A standard approach to inference integrates over the space of ranked labeled tree shapes generated by the Kingman coalescent \citep{drummond2012bayesian}. However, this inference is computationally intractable for large sample sizes. We have observed a striking reduction in the cardinality of the set of ranked unlabeled tree shapes compatible with an observed perfect phylogeny, relative to the number of ranked labeled tree shapes compatible with an observed perfect phylogeny (Table 1). This observation contributes to a growing branch of the area of coalescent-based inference \citep{veber, Palaciosgenetics, Palacios2019, Cappello2019} that can make use of ranked unlabeled trees to estimate the evolutionary parameters. 

%We have defined meet and join operations between perfect phylogenies of samples of size $n$, showing that the  poset $\Pi_{n}$ is a lattice. The lattice construction for the space of perfect phylogenies of size $n$ individuals allows us to define recursive expressions for calculating the number of lower resolution binary trees compatible with a given perfect phylogeny --- in particular, the number of ranked tree shapes and the number of tree shapes compatible with a given perfect phylogeny.

%{\bf XXX I AM NOT SURE I WOULD ADD THE FOLLOWING LAST SENTENCE XXX}
%We show that for a given perfect phylogeny, for example $(8,2)$ the number of ranked and labeled tree shapes compatible is $12,700,800$ while only $2,176$ ranked tree shapes, and only 23 tree shapes are compatible. 

\section{Acknowledgments}
J.A.P. and N.A.R. acknowledge support from National
Institutes of Health Grant R01-GM-131404. J.A.P. acknowledges support from the Alfred P. Sloan Foundation.
{\small
\bibliography{JP}
}

\section*{Appendix: Proof of Theorem \ref{thm:lattice}}

To prove Theorem \ref{thm:lattice}, we must verify four pairs of conditions concerning perfect phylogenies $\pi \in \Pi_{n} \cup \{\emptyset\}$. Note that any binary perfect phylogeny $\pi \in \Pi_{n} \cup \{\emptyset\}$ is equal to $\emptyset$, $(n)$, or $(\pi_1,\pi_2)$ for two non-empty binary perfect phylogenies $\pi_1 \in \Pi_{n_1}$ and $\pi_2 \in \Pi_{n_2}$, where $1 \leq n_1,n_2 < n$ and $n_1+n_2=n$. Hence, we must demonstrate the four pairs of conditions for perfect phylogeny pairs that include $\emptyset$, $(n)$, or both, and for perfect phylogeny pairs that include neither $\emptyset$ nor $(n)$. 

Because perfect phylogenies can be decomposed into smaller perfect phylogenies, we proceed by induction on $n$, with a base case of $n=1$. In the inductive step we assume that $(\Pi_{k} \cup \{\emptyset\}, \wedge, \vee)$ is a lattice for all $k$, $1 \leq k < n$. We then verify that it follows that $(\Pi_{n} \cup \{\emptyset\}, \wedge, \vee)$ is a lattice. We start with Condition 2, which is trivial.%; to prove the other three conditions, we first state and prove two lemmas. 

\subsection*{Condition 2: $\pi \wedge \sigma = \sigma \wedge \pi$ and $\pi \vee \sigma = \sigma \vee \pi$}
For all $n$, condition 2 of the definition of a lattice is trivially satisfied, as the operations $\wedge$ and $\vee$ are symmetric by definition. In subsequent derivations, we frequently apply Condition 2 without always noting its application.

\subsection*{The $n=1$ case for Conditions 1, 3, and 4}
Consider $n=1$, for which $\Pi_1$ contains only the perfect phylogeny $(1)$, and $\Pi_1 \cup \{\emptyset\}$ contains only $(1)$ and $\emptyset$. For $\Pi_1 \cup \{\emptyset\}$, demonstrating Condition 1 of the requirements for a lattice requires that we show $(1) \wedge (1) = (1)$, $\emptyset \wedge \emptyset = \emptyset$, $(1) \vee (1) = (1)$, and $\emptyset \vee \emptyset = \emptyset$. These four relations are true by parts (3), (1), (4), and (2) of Defn.~\ref{def:binope}, respectively.

Demonstrating Condition 3 requires that we verify a pair of conditions for each of the eight choices of $(x,y,z)$ for $x,y,z \in \Pi_1 \cup \{\emptyset\}$. Demonstrating Condition 4 requires that we verify a pair of conditions for each of the four choices of $(x,y)$. The 16 verifications for Condition 3 and eight verifications for Condition 4 all quickly follow by Defn.~\ref{def:binope} (1-4). Hence, $(\Pi_{1} \cup \{\emptyset\}, \wedge, \vee)$ is a lattice.

\subsection*{\bf Condition 1: $\pi \wedge \pi = \pi$ and $\pi \vee \pi = \pi$} First, we demonstrate the first part of the condition. We see $\emptyset \wedge \emptyset = \emptyset$ by Defn.~\ref{def:binope} (1) and $(n) \wedge (n) = (n)$ by Defn.~\ref{def:binope} (3). 

Consider $\pi = (\pi_1,\pi_2)$ for $\pi_1 \in \Pi_{n_1}$ and $\pi_2 \in \Pi_{n_2}$, where $1 \leq n_1,n_2 < n$ and $n_1 + n_2 = n$. 
\begin{align*}
    \pi \wedge \pi & = (\pi_1,\pi_2) \wedge (\pi_1,\pi_2) \\
    & = (\pi_1 \wedge \pi_1, \pi_2 \wedge \pi_2) \vee (\pi_1 \wedge \pi_2, \pi_2 \wedge \pi_1)  \text{ by Defn.~\ref{def:binope} (8)}\\
    & = (\pi_1, \pi_2) \vee  (\pi_1 \wedge \pi_2, \pi_1 \wedge \pi_2) \text{ by the inductive hypothesis}.
\end{align*}
If $n_1 \neq n_2$, then we apply Defn.~\ref{def:binope} (5), the convention $(\pi, \emptyset) = \emptyset$, and Defn.~\ref{def:binope} (2), and we obtain $\pi \wedge \pi = (\pi_1,\pi_2) \vee (\emptyset, \emptyset) = (\pi_1, \pi_2) \vee \emptyset = (\pi_1, \pi_2) = \pi$. If $n_1 = n_2$, then we have two cases: $\pi_1 \leq \pi_2$ (without loss of generality), and $\pi_1,\pi_2$ are not comparable.

If $\pi_1 \leq \pi_2$, then $\pi_1 \wedge \pi_2 = \pi_1$ and $\pi_1 \vee \pi_2 = \pi_2$. By Defn.~\ref{def:binope} (9), $(\pi_1,\pi_2) \vee (\pi_1,\pi_1) = (\pi_1, \pi_2 \vee \pi_1) = (\pi_1,\pi_2) = \pi$, so that $\pi \wedge \pi = \pi$.

If $\pi_1$ and $\pi_2$ are not comparable, then by Defn.~1 (11), $\pi_{1}\wedge \pi_{2}=\delta$ for some $\delta \in (\Pi_{n1}  \cup \{\emptyset\}) \setminus \{\pi_{1},\pi_{2}\}$, with $\delta \vee \pi_{1}=\pi_{1}$ and $\delta \vee \pi_{2}=\pi_{2}$. We then have by Defn.~\ref{def:binope} (9),
\begin{align*}
 (\pi_1, \pi_2) \vee  (\pi_1 \wedge \pi_2, \pi_1 \wedge \pi_2) & = 
(\pi_1, \pi_2)  \vee (\delta,\delta).
\end{align*}
But $(\delta,\delta)$ refines $(\pi_{1},\pi_{2})$, as $\delta$ refines $\pi_{1}$ and $\delta$ refines $\pi_{2}$, so that $(\pi_1, \pi_2)$ can be obtained by collapsing cherries separately in the two subtrees of $(\delta, \delta)$. Hence, $\pi \wedge \pi = (\pi_1, \pi_2) \vee (\delta, \delta) = (\pi_1, \pi_2) = \pi$.

For the second part of the condition, we have $\emptyset \vee \emptyset = \emptyset$ by Defn.~\ref{def:binope} (2) and $(n) \vee (n) = (n)$ by Defn.~\ref{def:binope} (4). Consider $\pi = (\pi_1,\pi_2)$ for $\pi_1 \in \Pi_{n_1}$ and $\pi_2 \in \Pi_{n_2}$, where $1 \leq n_1,n_2 < n$ and $n_1 + n_2 = n$. 
\begin{align*}
    \pi \vee \pi & = (\pi_1,\pi_2) \vee (\pi_1,\pi_2) \\
    & = (\pi_1, \pi_2 \vee \pi_2)  \text{ by Defn.~\ref{def:binope} (9)}\\
    & = (\pi_1, \pi_2) \text{ by the inductive hypothesis} \\
    & = \pi.
\end{align*}

\subsection*{Condition 4: $\pi \wedge (\pi \vee \sigma)=\pi$ and $\pi \vee (\pi \wedge \sigma)=\pi$} 
First, we see that both parts of the condition hold if at least one of $\pi, \sigma$ is in $\{ \emptyset,(n)\}$, by Defn.~1 (1-4). Next, we have the following 3 cases: 
\begin{enumerate}
    \item[i.] If $\pi \leq \sigma$, then  $\pi \wedge \sigma = \pi$ and $\pi \vee \sigma= \sigma$. Hence, $\pi \wedge (\pi \vee \sigma)=\pi \wedge \sigma = \pi$. By Condition 1, $\pi \vee (\pi \wedge \sigma)=\pi \vee \pi =\pi$. 
    \item[ii.] If $\sigma \leq \pi$, then $\pi \wedge \sigma =\sigma$ and $\pi \vee \sigma=\pi$. Hence, by Condition 1, $\pi \wedge (\pi \vee \sigma)=\pi \wedge \pi = \pi$. We also have $\pi \vee (\pi \wedge \sigma)=\pi \vee \sigma =\pi$. 
    \item[iii.] If $\pi$ and $\sigma$ are not comparable, then by Defn.~1 (11), there exists a perfect phylogeny $\gamma$ such that $\pi \vee \sigma = \gamma$, $\pi \wedge \gamma=\pi$, and $\sigma \wedge \gamma=\sigma$.  Hence $\pi \wedge(\pi \vee \sigma)=\pi \wedge \gamma = \pi$. By Defn.~1 (11), there exists a perfect phylogeny $\rho$ such that $\pi \wedge \sigma=\rho$, $\pi \vee \rho=\pi$, and $\sigma \vee \rho = \sigma$. We have $\pi \vee (\pi \wedge \sigma)=\pi \vee \rho =\pi$. 
\end{enumerate}

\subsection*{Condition 3: $\pi \wedge (\sigma \wedge  \rho)=(\pi \wedge \sigma) \wedge  \rho$ and $\pi \vee (\sigma \vee  \rho)=(\pi \vee \sigma) \vee  \rho$} 
First, we see that both parts of the condition hold if at least one of $\pi, \sigma, \rho$ is in $\{\emptyset,(n)\}$, by Defn.~1 (1-4). Assume now that $\pi=(\pi_{1},\pi_{2})$, $\sigma=(\sigma_{1},\sigma_{2})$, and  $\rho=(\rho_{1},\rho_{2})$. Then
 \begin{align*}
\pi \wedge (\sigma \wedge  \rho) &=(\pi_{1},\pi_{2}) \wedge   \left( (\sigma_{1},\sigma_{2}) \wedge  (\rho_{1},\rho_{2}) \right)\\
&=(\pi_{1},\pi_{2}) \wedge [ (\sigma_{1} \wedge \rho_{1}, \sigma_{2}\wedge \rho_{2}) \vee (\sigma_{1} \wedge \rho_{2}, \sigma_{2}\wedge \rho_{1})] \text{ by Defn.~\ref{def:binope} (8)}\\
&=[(\pi_{1},\pi_{2})\wedge (\sigma_{1}\wedge \rho_{1},\sigma_{2}\wedge\rho_{2})]\vee [(\pi_{1},\pi_{2})\wedge (\sigma_{1}\wedge\rho_{2},\sigma_{2}\wedge \rho_{1})] \text{
by Defn.~\ref{def:binope} (10)}\\
&=[(\pi_{1}\wedge (\sigma_{1} \wedge \rho_{1}), \pi_{2} \wedge (\sigma_{2}\wedge \rho_{2}))\vee(\pi_{1}\wedge (\sigma_{2} \wedge \rho_{2}), \pi_{2} \wedge (\sigma_{1}\wedge \rho_{1})) ] \\
& \quad \vee [(\pi_{1}\wedge (\sigma_{1} \wedge \rho_{2}), \pi_{2} \wedge (\sigma_{2}\wedge \rho_{1}))\vee(\pi_{1}\wedge (\sigma_{2} \wedge \rho_{1}), \pi_{2} \wedge (\sigma_{1}\wedge \rho_{2}) )]\text{ by Defn.~\ref{def:binope} (8)}
\end{align*}
By the inductive hypothesis {\it for both parts of the condition}, $\pi_{i} \wedge (\sigma_{j} \wedge \rho_{k})= (\pi_{i} \wedge \sigma_{j})\wedge \rho_{k}$ and $\pi_{i} \vee (\sigma_{j} \vee \rho_{k})= (\pi_{i} \vee \sigma_{j})\vee \rho_{k}$
for all $i,j,k \in \{1,2 \}$. We then get
 \begin{align*}
\pi \wedge (\sigma \wedge  \rho) &= [((\pi_{1}\wedge \sigma_{1}) \wedge \rho_{1}, (\pi_{2} \wedge \sigma_{2})\wedge \rho_{2})\vee ((\pi_{1}\wedge \sigma_{2}) \wedge \rho_{2}, (\pi_{2} \wedge \sigma_{1})\wedge \rho_{1})]\\  
& \quad \vee [((\pi_{1}\wedge \sigma_{1}) \wedge \rho_{2}, (\pi_{2} \wedge \sigma_{2})\wedge \rho_{1})\vee ((\pi_{1}\wedge \sigma_{2}) \wedge \rho_{1}, (\pi_{2} \wedge \sigma_{1})\wedge \rho_{2})].
\end{align*}
By the inductive hypothesis for operator $\vee$ and by Condition 2, we can rearrange parentheses and swap the order of terms to obtain:
 \begin{align*}
\pi \wedge (\sigma \wedge  \rho) &=((\pi_{1}\wedge \sigma_{1}) \wedge \rho_{1}, (\pi_{2} \wedge \sigma_{2})\wedge \rho_{2})\vee [(\pi_{1}\wedge \sigma_{1}) \wedge \rho_{2}, (\pi_{2} \wedge \sigma_{2})\wedge \rho_{1})\\  
& \quad \vee ((\pi_{1}\wedge \sigma_{2}) \wedge \rho_{2}, (\pi_{2} \wedge \sigma_{1})\wedge \rho_{1}]\vee ((\pi_{1}\wedge \sigma_{2}) \wedge \rho_{1}, (\pi_{2} \wedge \sigma_{1})\wedge \rho_{2}).
\end{align*}
Dropping the brackets and viewing this expression as having four perfect phylogenies separated by the $\vee$ operator, we group the first two and the last two perfect phylogenies together and apply Defn.~1 (8) to each group. We get
\begin{align*}
\pi \wedge (\sigma \wedge  \rho) &=[(\pi_{1}\wedge \sigma_{1},\pi_{2}\wedge \sigma_{2}) \wedge (\rho_{1},\rho_{2})] \vee [(\pi_{1}\wedge \sigma_{2},\pi_{2}\wedge \sigma_{1}) \wedge (\rho_{1},\rho_{2})] \text{ by Defn.~1 (8)}\\
&=[(\pi_{1}\wedge \sigma_{1},\pi_{2}\wedge \sigma_{2}) \vee (\pi_{1}\wedge \sigma_{2},\pi_{2}\wedge \sigma_{1})] \wedge (\rho_{1},\rho_{2}) \text{ by Defn.~1 (10)}\\
&=(\pi\wedge \sigma)\wedge \rho \text{ by Defn.~1 (8).}
\end{align*}

For the second part of the condition, suppose $\pi=(\pi_{1},\pi_{2}) \in \Pi_{n}$,
$\sigma=(\sigma_{1},\sigma_{2}) \in \Pi_{n}$, and $\rho=(\rho_{1},\rho_{2})\in \Pi_{n}$ are three perfect phylogenies of size $n$. We consider four cases. First, suppose the three perfect phylogenies have mutually different subtree sizes---that is, $\{|\pi_1|,|\pi_2|\}$, $\{|\sigma_1|,|\sigma_2|\}$, and $\{|\rho_1|,|\rho_2|\}$ are mutually distinct. Then $\pi \vee \sigma = \pi \vee \rho = \sigma \vee \rho=(n)$ by Defn.~\ref{def:binope}~(9). We then have $\pi \vee (\sigma \vee \rho)=\pi \vee (n)=(n)= (n) \vee \rho = (\pi\vee \sigma) \vee \rho$ by Defn.~1 (4). 

%the subtrees $\sigma_{1},\sigma_{2}$ have sizes different from $\rho_{1},\rho_{2}$ \textcolor{blue}{(more precisely, if $\{|\sigma_1|,|\sigma_2|\} \neq \{|\rho_1|,|\rho_2|\}$)},  then $\sigma \vee \rho=(n)$ \textcolor{blue}{by Defn.~\ref{def:binope}~(7)}. Therefore, if all pairs of subtrees of $\pi$, $\sigma$, and $\rho$ have different sizes \textcolor{blue}{($\{|\pi_1|,|\pi_2|\}$, $\{|\sigma_1|,|\sigma_2|\}$, and $\{|\rho_1|,|\rho_2|\}$ mutually distinct)}, then $\pi \vee (\sigma \vee \rho)=\pi \vee (n)=(n)=(\pi\vee \sigma) \vee \rho$ by Defn.~1 (4,9). 

The same argument applies if it is merely assumed that $\sigma$ and $\rho$ have pairs of subtrees whose sizes differ, $\{|\sigma_1|,|\sigma_2|\} \neq \{|\rho_1|,|\rho_2|\})$. Then $\pi \vee (\sigma \vee \rho)=\pi\vee(n)=(n)=(\pi\vee\sigma)\vee \rho$ by Defn.~1 (4, 9), where we have used the fact that $\sigma \vee \rho = (n)$ and $\sigma \leq \pi \vee \sigma$, so that $(\pi \vee \sigma) \vee \rho = (n)$.

If $\{|\sigma_1|,|\sigma_2|\} = \{|\rho_1|,|\rho_2|\})$ but $\{|\pi_1|,|\pi_2|\} \neq \{|\sigma_1|,|\sigma_2|\})$ and $\{|\pi_1|,|\pi_2|\} \neq \{|\rho_1|,|\rho_2|\})$, then $\pi \vee \sigma = (n)$. Because $\sigma \leq \sigma \vee \rho$ and $\pi \vee \sigma = (n)$, $\pi \vee (\sigma \vee \rho) = (n)$. Similarly, $(\pi\vee\sigma)\vee \rho = (n) \vee \rho = (n)$ by Defn.~1 (4, 9).

It remains to consider the case in which at least a pair of subtrees, one each from $\pi$, $\sigma$ and $\rho$ have the same size, or $\{|\pi_1|,|\pi_2|\} = \{|\sigma_1|,|\sigma_2|\}) = \{|\rho_1|,|\rho_2|\})$. We have
\begin{align}
\pi \vee (\sigma \vee  \rho) &=(\pi_{1},\pi_{2}) \vee   \left( (\sigma_{1},\sigma_{2}) \vee  (\rho_{1},\rho_{2}) \right) \nonumber \\
&=(\pi_{1},\pi_{2}) \vee [ (\sigma_{1} \vee \rho_{1}, \sigma_{2}\vee \rho_{2}) \wedge (\sigma_{1} \vee \rho_{2}, \sigma_{2}\vee \rho_{1})] \text{ by Defn.~\ref{def:binope} (9)} \nonumber \\
&=[(\pi_{1},\pi_{2})\vee (\sigma_{1}\vee \rho_{1},\sigma_{2}\vee\rho_{2})]\wedge [(\pi_{1},\pi_{2})\vee (\sigma_{1}\vee \rho_{2},\sigma_{2}\vee \rho_{1})] \text{
by Defn.~\ref{def:binope} (10)} \nonumber \\
&=[(\pi_{1}\vee (\sigma_{1} \vee \rho_{1}), \pi_{2} \vee (\sigma_{2}\vee \rho_{2}))\wedge(\pi_{1}\vee (\sigma_{2} \vee \rho_{2}), \pi_{2} \vee (\sigma_{1}\vee \rho_{1})) ] \nonumber \\
&\quad \wedge [(\pi_{1}\vee (\sigma_{1} \vee \rho_{2}), \pi_{2} \vee (\sigma_{2}\vee \rho_{1}))\wedge(\pi_{1}\vee (\sigma_{2} \vee \rho_{1}), \pi_{2} \vee (\sigma_{1}\vee \rho_{2})) ]\text{ by Defn.~\ref{def:binope} (9)} \nonumber \\
&=((\pi_{1}\vee \sigma_{1}) \vee \rho_{1}, (\pi_{2} \vee \sigma_{2})\vee \rho_{2})\wedge ((\pi_{1}\vee \sigma_{1}) \vee \rho_{2}, (\pi_{2} \vee \sigma_{2})\vee \rho_{1}) \nonumber \\  
&\quad \wedge((\pi_{1}\vee \sigma_{2}) \vee \rho_{2}, (\pi_{2} \vee \sigma_{1})\vee \rho_{1})\wedge ((\pi_{1}\vee \sigma_{2}) \vee \rho_{1}, (\pi_{2} \vee \sigma_{1})\vee \rho_{2}) \text{ by ind.~hypothesis} \nonumber \\
&=[(\pi_{1}\vee \sigma_{1},\pi_{2}\vee \sigma_{2}) \vee (\rho_{1},\rho_{2})] \wedge [(\pi_{1}\vee \sigma_{2},\pi_{2}\vee \sigma_{1}) \vee (\rho_{1},\rho_{2})] \text{ by Defn.~1 (9)} \nonumber \\
&=[(\pi_{1}\vee \sigma_{1},\pi_{2}\vee \sigma_{2}) \wedge (\pi_{1}\vee \sigma_{2},\pi_{2}\vee \sigma_{1})] \vee (\rho_{1},\rho_{2}) \text{ by Defn.~1 (10)}\nonumber \\
&=(\pi\vee \sigma)\vee \rho \text{ by Defn.~1 (9).}
\label{eq20}
\end{align}
Note that this derivation includes the case of shared subtrees at the root, in which it is not only the sizes of the subtrees that are the same, but the subtrees themselves. For example, suppose $\pi=(\pi_{1},\pi_{2})$ and $\sigma=(\pi_{1},\sigma_{1})$. By Defn.~1 (9), we have
\begin{align*}
    \pi\vee\sigma &= (\pi_{1},\pi_{2})\vee (\pi_{1},\sigma_{1})= (\pi_{1},\pi_{2}\vee \sigma_{1}).
\end{align*}
However, we will show that we can replace the previous equality by the extended expression:
\begin{align}
\label{eq21}
    \pi\vee\sigma &= (\pi_{1}\vee \pi_{1},\pi_{2}\vee\sigma_{1}) \wedge (\pi_{1}\vee \sigma_{1},\pi_{1}\vee \pi_{2}),
\end{align}
and then the previous derivation remains unchanged. To prove this assertion, we have:
\begin{align*}
    \pi_{2}\vee \sigma_{1}&=(\pi_{2} \vee (\pi_{1}\wedge \pi_{2})) \vee \sigma_{1} \quad \text{ by Condition 4}\\
    &=(\pi_{2} \vee \sigma_{1}) \vee (\pi_{1}\wedge \pi_{2}) 
    \quad \text{ by the inductive hypothesis and Condition 2}\\
    &=\pi_{2}\vee [(\sigma_{1} \wedge \pi_{1})\vee \sigma_{1} ]\vee (\pi_{1}\wedge \pi_{2}) \quad \text{ by Conditions 2 and 4}\\
    &=[\pi_{2}\vee (\sigma_{1} \wedge \pi_{1})]\vee [\sigma_{1}\vee (\pi_{1}\wedge \pi_{2})] \quad \text{ by the inductive hypothesis.}
\end{align*}
Then
\begin{align}\label{eq22}
    \pi\vee\sigma &= (\pi_{1},\pi_{2}\vee \sigma_{1}) \nonumber \\
    &=(\pi_{1},[\pi_{2}\vee (\sigma_{1} \wedge \pi_{1})]\vee [\sigma_{1}\vee (\pi_{1}\wedge \pi_{2})]) \nonumber \\
    &=(\pi_{1},\pi_{2}\vee (\sigma_{1}\wedge \pi_{1}))\vee (\pi_{1},\sigma_{1}\vee(\pi_{1}\wedge \pi_{2})) \quad \text{ by Defn.~1 (9)} \nonumber \\
    &=(\pi_{1},(\pi_{2}\vee \sigma_{1}) \wedge (\pi_{2}\vee \pi_{1})) \vee (\pi_{1},(\sigma_{1}\vee\pi_{1})\wedge (\sigma_{1}\vee\pi_{2})) \quad \text{ by Defn.~1 (10).}
\end{align}
By Condition 4 and Defn.~1 (9) we have
\begin{align*}
\pi_{1}&=\pi_{1} \vee (\pi_{1}\wedge\sigma_{1})=(\pi_{1} \vee \pi_{1}) \wedge ( \pi_{1}\vee\sigma_{1}),
\end{align*}
and
\begin{align*}
\pi_{1}&=\pi_{1} \vee (\pi_{1}\wedge \pi_{2})= (\pi_{1}  \vee \pi_{1})\wedge (\pi_{1}\vee \pi_{2}).
\end{align*}
Replacing the first $\pi_{1}$ in the first pair of eq.~\ref{eq22} by $(\pi_{1} \vee \pi_{1}) \wedge ( \pi_{1}\vee\sigma_{1})$, and the first $\pi_{1}$ in the second pair of eq.~\ref{eq22} by $(\pi_{1} \vee \pi_{1}) \wedge ( \pi_{1}\vee\pi_{2})$, we get
\begin{align*}
    \pi\vee\sigma &=((\pi_{1} \vee \pi_{1}) \wedge ( \pi_{1}\vee\sigma_{1}),(\pi_{2}\vee \sigma_{1}) \wedge (\pi_{2}\vee \pi_{1})) \vee ((\pi_{1} \vee \pi_{1}) \wedge ( \pi_{1}\vee\pi_{2}),(\sigma_{1}\vee\pi_{1})\wedge (\sigma_{1}\vee\pi_{2}))\\
    &=(\pi_{1}\vee \pi_{1},\pi_{2}\vee\sigma_{1}) \wedge (\pi_{1}\vee \sigma_{1},\pi_{1}\vee \pi_{2})\quad \text{ by Defn.~1 (8).}
    \end{align*}
Thus, eq.~\ref{eq21} holds, so that eq.~\ref{eq20} holds for the case in which subtrees are shared at the root.

\clearpage

\end{document}